\documentclass[12pt]{article}
\usepackage{a4wide,epsfig,amsmath,amssymb,cite,scalefnt,graphicx,epstopdf}
\numberwithin{equation}{section}
\usepackage{axodraw4j}
\usepackage{pstricks}
\usepackage{color}
\usepackage{changepage}
\usepackage{tabularx}
\usepackage{graphics}
\usepackage{array}

\def\npb{{\it{Nucl.\ Phys.}\ }{\bf B}}

\def\be{\begin{equation}}
\def\ee{\end{equation}}
\def\bea{\begin{align}}
\def\eea{\end{align}}
\def\nn{\nonumber\\}

\def\tr{\hbox{tr}}

\def\msbar{{\overline{\rm MS}}}

\def\psibar{\overline{\psi}}

\def\frakk[#1#2{{{#1}\over{#2}}}

\def\psibar{\overline\psi}

\def\Ncal{{\cal N}}

\def\pa{\partial}

\input amssym.def
\input amssym
\def\npb{Nucl. Phys. B}

\def\pa{\partial}
\def\be{\begin{equation}}
\def\ee{\end{equation}}
\def\bea{\begin{align}}
\def\eea{\end{align}}
\def\nn{\nonumber\\}

\def\tr{\hbox{tr}}

\def \tr{{\rm tr }}

\def\cirk{\,{\raise1pt \hbox{${\scriptscriptstyle \circ}$}}\,}

\def \olr{{\raise6.5pt\hbox{$\leftrightarrow  \! \! \! \! \!$}}}

\font\ninerm=cmr9 \font\ninesy=cmsy9
\font\eightrm=cmr8 \font\sixrm=cmr6
\font\eighti=cmmi8 \font\sixi=cmmi6
\font\eightsy=cmsy8 \font\sixsy=cmsy6
\font\eightbf=cmbx8 \font\sixbf=cmbx6
\font\eightit=cmti8
\def\eightpoint{\def\rm{\fam0\eightrm}
  \textfont0=\eightrm \scriptfont0=\sixrm \scriptscriptfont0=\fiverm
  \textfont1=\eighti  \scriptfont1=\sixi  \scriptscriptfont1=\fivei
  \textfont2=\eightsy \scriptfont2=\sixsy \scriptscriptfont2=\fivesy
  \textfont3=\tenex   \scriptfont3=\tenex \scriptscriptfont3=\tenex
  \textfont\itfam=\eightit  \def\it{\fam\itfam\eightit}%
  \textfont\bffam=\eightbf  \scriptfont\bffam=\sixbf
  \scriptscriptfont\bffam=\fivebf  \def\bf{\fam\bffam\eightbf}%
  \normalbaselineskip=9pt
  \setbox\strutbox=\hbox{\vrule height7pt depth2pt width0pt}%
  \let\big=\eightbig  \normalbaselines\rm}
\catcode`@=11 %
\def\eightbig#1{{\hbox{$\textfont0=\ninerm\textfont2=\ninesy
  \left#1\vbox to6.5pt{}\right.\n@@space$}}}
\def\vfootnote#1{\insert\footins\bgroup\eightpoint
  \interlinepenalty=\interfootnotelinepenalty
  \splittopskip=\ht\strutbox %
  \splitmaxdepth=\dp\strutbox %
  \leftskip=0pt \rightskip=0pt \spaceskip=0pt \xspaceskip=0pt
  \textindent{#1}\footstrut\futurelet\next\fo@t}
\catcode`@=12 %
\def\today{\number\day\ \ifcase\month\or January\or February\or March\or
April\or May\or June\or July\or
August\or September\or October\or November\or December\fi, \number\year}

\input epsf

\begin{document}

\begin{titlepage}
\begin{flushright}
LTH1082\\
\end{flushright}
\date{}
\vspace*{3mm}

\begin{center}
{\Huge The $a$-function in three dimensions:\\ beyond leading order}\\[12mm]
{\bf I.~Jack\footnote{{\tt dij@liv.ac.uk}} and
C.~Poole\footnote{{\tt c.poole@liv.ac.uk}}}\\

\vspace{5mm}
Dept. of Mathematical Sciences,
University of Liverpool, Liverpool L69 3BX, UK\\

\end{center}

\vspace{3mm}
\begin{abstract}
Recently, evidence was provided for the existence of an $a$-function for renormalisable quantum field theories in three dimensions. An explicit expression was given at lowest order for general theories involving scalars and fermions, and shown to be related to the $\beta$-functions by a gradient flow equation with positive-definite metric as in four dimensions. 
Here, we extend this lowest-order calculation to a general abelian Chern-Simons gauge theory coupled to fermions and scalars, and derive a prediction for part of the four-loop Yukawa $\beta$-function. We also compute the complete four-loop Yukawa $\beta$-function for the scalar-fermion theory and show that it is entirely consistent with the gradient flow equations at next-to-leading order.
\end{abstract}

\vfill

\end{titlepage}

\section{Introduction}

Following Cardy's suggestion\cite{Cardy} that Zamolodchikov's two-dimensional $c$-theorem\cite{Zam} might have an analogue in four dimensions, considerable progress has been made in proving the so-called $a$-theorem in even dimensions \cite{KS,Luty,ElvangST,Analog,OsbJacnew,Weyl,GrinsteinCKA}. In a recent paper\cite{JJP} we provided evidence that for a wide range of renormalisable quantum field theories in three dimensions
we can similarly define a function $A$ which satisfies the equation
\be \pa_I A =T_{IJ}\beta^J\, ,
\label{grad}
\ee
for a function  $T_{IJ}$; we denote the function by $A$ since the notation $a$ is often used in four dimensions for the Euler density coefficient in the Weyl anomaly. A crucial consequence of Eq.~\eqref{grad} is that we then have
\be
\mu \frac{d}{d\mu} A=\beta^I\frac{\pa}{\pa g^I} A=G_{IJ} \beta^I\beta^J
\ee
where $G_{IJ}=T_{(IJ)}$; thus demonstrating a function with monotonic behaviour under
renormalisation group (RG) flow and providing a three-dimensional version of the strong $a$-theorem so long as $G_{IJ}$ is positive-definite. This is remarkable, since attempts\cite{Nakthree} to extend the methods\cite{Analog} used to prove the strong $a$-theorem to three dimensions did not appear to lead to a relation of the desired form. 
In Ref.~\cite{JJP} we firstly used the leading-order (two-loop) $\beta$-functions computed in Refs.~\cite{kaza,kazb} to construct  a solution of Eq.~\eqref{grad} for Abelian and non-Abelian (for the case $SU(n)$) Chern-Simons theories at leading order. Our method was essentially that employed in four dimensions in the classic work of Ref.~\cite{Wallace}. The ``metric'' $G_{IJ}$ was indeed found to be positive definite at this order, at least perturbatively. 
The Yukawa and scalar couplings in these theories were of a restricted form. However, by considering completely general scalar/fermion theories (but without gauge interactions)  we were able to argue that the existence of the $a$-function was somewhat trivial for these theories at leading order; but that predictions for the scalar-coupling-dependent contributions to the next-to-leading order (four-loop) Yukawa $\beta$-function emerged and could be verified by an explicit computation.

 In this paper our purpose is first of all to extend the general leading order calculation to the gauged case (we present results for the abelian case, but the extension to the non-abelian case is straightforward); and secondly to complete the four-loop computation\footnote{modulo the anomalous dimensions for which we do not have a fully independent computation, as we shall explain later.} of the Yukawa $\beta$-function \cite{JJP} for a general scalar/fermion theory and show that we can extend the definition of the $a$-function in Eq.~\eqref{grad} to this order. It turns out that in the gauged case, the existence of the $a$-function is non-trivial even at leading order; it imposes constraints on the $\beta$-function coefficients which we will show are satisfied. A by-product of our extended leading-order computation is a prediction for the scalar-coupling-dependent contribution to the four-loop Yukawa $\beta$-function for a completely general (i.e. gauged) renormalisable theory in three dimensions. 

It has already been
proposed that the free energy $F$ in three dimensions  may have similar
properties to the four-dimensional $a$-function, leading to a conjectured
``$F$-theorem''\cite{jaff,klebb,kleba,klebc}. It has been shown that for
certain theories in three  dimensions, the free energy does indeed
decrease monotonically along RG trajectories.
It has also been shown that $F$ obeys a gradient flow equation at leading order for theories which may be regarded as a perturbation around a conformal field theory. This covers theories with a scalar potential, which may be regarded a perturbation around a free field theory. However the only non-trivial example at leading order, namely a gauged scalar-fermion theory, does not fall into this class. This is the reason why we have pursued the computation beyond leading order despite its complexities. On the other hand, our method does not provide any general insight as to the origin of the gradient flow, so it would be interesting to investigate the relation between 
the ``$F$-function'' and our $a$-function.

The structure of the paper is as follows. In Section 2 we discuss the construction of the $a$-function at leading order, corresponding to the two-loop Yukawa $\beta$-function. Here we consider a completely general Chern-Simons gauge theory coupled to fermions and scalars, and we show that in this gauged case, Eq.~\eqref{grad} imposes non-trivial constraints on the $\beta$-function coefficients which are indeed satisfied. In Section 3 we proceed to the next-to-leading order, but for a general ungauged scalar/fermion theory. Here the $a$-function is determined by the two-loop scalar $\beta$-function and the four-loop Yukawa $\beta$-function.  We show that Eq.~\eqref{grad} imposes a plethora of constraints upon the  four-loop Yukawa $\beta$-function coefficients; and we compute the four-loop $\beta$-function to demonstrate that these are all satisfied. Various remarks are offered in a conclusion. Finally, a number of technical details are postponed to appendices: namely, an explicit list of the tensor structures in which we expressed the $\beta$-function results, together with  an explanation of our choice of these structures; and the full set of consistency conditions and four-loop Yukawa $\beta$-function results at next-to-leading order. We also discuss there the scheme dependence of our results. We present in a final appendix our prediction for the scalar-coupling-dependent contribution to the general four-loop Yukawa $\beta$-function, as mentioned above.

\section{Leading order results}

In this section we define the general three-dimensional abelian Chern-Simons theory, present its $\beta$-functions at lowest order (two loops) and construct the leading term in the $a$-function. The lagrangian is given by
\begin{align}
L=&\tfrac12[\epsilon^{\mu\nu\rho}A_{\mu}\pa_{\nu}A_{\rho}+(D_{\mu}\phi_i)^2+i\psibar_a{ D}\psi_a]\nn
&+\tfrac14 Y_{abij}\psi_a\psi_b\phi_i\phi_j
-\tfrac{1}{6!}h_{ijklmn}\phi_i\phi_j\phi_k\phi_l\phi_m\phi_n
\label{lagc}
\end{align}
where  we employ a real basis for both scalar and fermion fields, and $D_{\mu}=\pa_{\mu}-iEA_{\mu}$ where $E$ is a charge matrix ($E^{\phi}$,  $E^{\psi}$, for scalar, fermion fields respectively)
Recall that in $d=3$, $\psibar = \psi^{*T}$, and there is  no obstacle to
decomposing $\psi$  into real Majorana fields. Gauge invariance entails the identities
\begin{align}
E^{\psi}_{ac}Y_{cbij}+E^{\psi}_{bc}Y_{acij}+E^{\phi}_{im}Y_{abmj}+E^{\phi}_{jm}Y_{abim}=&0,\nn
E_{ip}^{\phi}h_{pjklmn}+\hbox{perms}=&0.
\label{gauge}
\end{align}

The $L$-loop Yukawa and scalar $\beta$-functions take the respective forms
\begin{equation}
(\beta^{(L)}_Y)_{abij} = \sum\limits_{\alpha=1}^{n_L} c^{(L)}_{\alpha}({U}_{\alpha}^{(L)})_{abij}, \;\;\;\;  (\beta^{(L)}_h)_{ijklmn} = \sum\limits_{\alpha=1}^{m_L} d^{(L)}_{\alpha}({V}_{\alpha}^{(L)})_{ijklmn},
\label{betdef}
\end{equation}
where ${U}^{(L)}_{\alpha}$, ${V}^{(L)}_{\alpha}$  denote $L$-loop tensor structures. In the interests of brevity, in the main body of the text we shall simply give a diagrammatic representation of the various tensor stuctures appearing here; to avoid any ambiguity the full expressions will be given in Appendix \ref{A1}.  In these diagrams the Yukawa and scalar couplings will be represented by vertices, with the fermion and scalar legs indicated thus, 
with lines indicating contracted indices:
 
 \begin{center}
 	\vspace{-2cm}
 	$Y_{abij}$ $\rightarrow$ \scalebox{0.45}
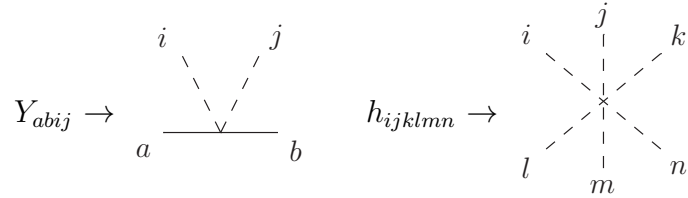
{\begin{picture}(160,142) (145,-120)
 		\SetWidth{1.0}
 		\SetColor{Black}
 		\Line(176,-128)(272,-128)
 		\Line[dash,dashsize=10](224,-128)(192,-64)
 		\Line[dash,dashsize=10](224,-128)(256,-64)
 		\Text(176,-48)[]{\Huge{\Black{$i$}}}
 		\Text(288,-144)[]{\Huge{\Black{$b$}}}
 		\Text(160,-144)[]{\Huge{\Black{$a$}}}
 		\Text(272,-48)[]{\Huge{\Black{$j$}}}
 		\end{picture}
 	}$\;\;\;\;$
 	$h_{ijklmn}$ $\rightarrow$ \scalebox{0.45}{\begin{picture}(160,220) (161,-74)
 		\SetWidth{1.0}
 		\SetColor{Black}
 		\Line[dash,dashsize=10](240,0)(240,-112)
 		\Line[dash,dashsize=10](192,-16)(288,-96)
 		\Line[dash,dashsize=10](288,-16)(192,-96)
 		\Text(176,0)[]{\Huge{\Black{$i$}}}
 		\Text(304,-112)[]{\Huge{\Black{$n$}}}
 		\Text(240,-128)[]{\Huge{\Black{$m$}}}
 		\Text(176,-112)[]{\Huge{\Black{$l$}}}
 		\Text(304,0)[]{\Huge{\Black{$k$}}}
 		\Text(240,16)[]{\Huge{\Black{$j$}}}
 		\end{picture}
 	}
 \vspace{0.5cm}
 \end{center}

At two loops, the number of tensor structures appearing in the Yukawa $\beta$-function in Eq.~\eqref{betdef} is given by  
$n_2=29$, and the two-loop tensor structures are displayed in Table~\ref{fig3} and written explicitly in Eqs.~\eqref{Ynone}, \eqref{Ytwo} and \eqref{Yfour}. A small circle represents a single gauge matrix $E^{\phi}$ or $E^{\psi}$, and a square represents a product of two $E^{\phi}$ or $E^{\psi}$.  Each tensor structure is defined so as to have a ``weight" of one, as explained in Appendix \ref{A1}; where we also explain our choice for these structures, which is not unique since tensor structures containing gauge matrices may be
related through the gauge-invariance identity Eq.~\eqref{gauge}.    
We note here that $U^{(2)}_{22}$-$U^{(2)}_{25}$  correspond to anomalous dimension contributions and consequently we may simply read off the corresponding values of $c^{(2)}_{22}$-$c^{(2)}_{25}$ from the results of Ref.~\cite{kaza} with no further calculation. In the case of $U_{24}^{(2)}$ and $U_{25}^{(2)}$, there is also a graph with a fermion loop which is not depicted but whose contribution may be seen in Eq.~\eqref{Yfour}.We have assumed that the contributions from single fermion loops and single scalar loops are equal; this is consistent with our other explicit calculations but in any case does not affect any of our conclusions.

\begin{table}[t]
	\setlength{\extrarowheight}{1cm}
	\setlength{\tabcolsep}{24pt}
	\hspace*{-7.75cm}
	\centering
	\resizebox{6.7cm}{!}{
		\begin{tabular*}{20cm}{ccccc}
			\begin{picture}(162,87) (303,-240)
			\SetWidth{1.0}
			\SetColor{Black}
			\Line(304,-201)(464,-201)
			\Arc[dash,dashsize=9.6](392,-179.382)(60.028,-158.891,-21.109)
			\Arc[dash,dashsize=9](416,-192.988)(32.988,-14.057,194.057)
			\Line[dash,dashsize=10](336,-201)(320,-153)
			\Line[dash,dashsize=10](384,-201)(368,-153)
			\end{picture}
			&
			\begin{picture}(162,150) (303,-191)
			\SetWidth{1.0}
			\SetColor{Black}
			\Line(304,-138)(464,-138)
			\Line[dash,dashsize=10](336,-42)(384,-138)
			\Line[dash,dashsize=10](432,-42)(384,-138)
			\Arc[dash,dashsize=9.6](384,-118)(52,-157.38,-22.62)
			\Arc[dash,dashsize=10.6](384,-141.846)(48.154,175.419,364.581)
			\end{picture}
			&
			\begin{picture}(162,98) (303,-191)
			\SetWidth{1.0}
			\SetColor{Black}
			\Line(304,-190)(464,-190)
			\Line[dash,dashsize=8.2](336,-94)(384,-190)
			\Line[dash,dashsize=8.2](432,-94)(384,-190)
			\Arc[clock](384,-150)(40,143.13,36.87)
			\Arc(384,-102)(40,-143.13,-36.87)
			\end{picture}
			&
			\begin{picture}(162,128) (303,-191)
			\SetWidth{1.0}
			\SetColor{Black}
			\Line(304,-160)(464,-160)
			\Line[dash,dashsize=8.2](304,-64)(352,-160)
			\Line[dash,dashsize=8.2](400,-64)(352,-160)
			\Arc[dash,dashsize=9.4](416,-160)(30.017,150,510)
			\end{picture}
			&
			\begin{picture}(162,98) (303,-191)
			\SetWidth{1.0}
			\SetColor{Black}
			\Line(304,-190)(464,-190)
			\Line[dash,dashsize=8.2](336,-94)(384,-190)
			\Line[dash,dashsize=8.2](432,-94)(384,-190)
			\Arc(412,-134)(22.627,135,495)
			\end{picture}
			\\
			{\Huge $(U^{(2)}_{1})_{abij}$}
			&
			{\Huge $(U^{(2)}_{2})_{abij}$}
			&
			{\Huge $(U^{(2)}_{3})_{abij}$}
			&
			{\Huge $(U^{(2)}_{4})_{abij}$}
			&
			{\Huge $(U^{(2)}_{5})_{abij}$}
			\\
			&
			&
			&
			&
			\\
			\begin{picture}(162,78) (303,-223)
			\SetWidth{1.0}
			\SetColor{Black}
			\Line(304,-210)(464,-210)
			\Arc[dash,dashsize=10,clock](384.5,-214.669)(63.671,175.794,4.206)
			\Line[dash,dashsize=10](321,-210)(304,-146)
			\Line[dash,dashsize=10](449,-210)(464,-146)
			\CBox(396,-222)(372,-198){Black}{Black}
			\end{picture}
			&
			\begin{picture}(162,76) (303,-221)
			\SetWidth{1.0}
			\SetColor{Black}
			\Line(304,-212)(464,-212)
			\Arc[dash,dashsize=10,clock](384.5,-216.669)(63.671,175.794,4.206)
			\Line[dash,dashsize=10](321,-212)(304,-148)
			\Line[dash,dashsize=10](449,-212)(464,-148)
			\Vertex(384,-154){8}
			\Vertex(384,-212){8}
			\end{picture}
			&
			\begin{picture}(194,74) (287,-223)
			\SetWidth{1.0}
			\SetColor{Black}
			\Line(288,-214)(480,-214)
			\Arc[dash,dashsize=10,clock](384.5,-218.669)(63.671,175.794,4.206)
			\Line[dash,dashsize=10](321,-214)(304,-150)
			\Line[dash,dashsize=10](449,-214)(464,-150)
			\Vertex(464,-214){8}
			\Vertex(304,-214){8}
			\end{picture}
			&
			\begin{picture}(194,76) (287,-221)
			\SetWidth{1.0}
			\SetColor{Black}
			\Line(288,-212)(480,-212)
			\Arc[dash,dashsize=10,clock](384.5,-216.669)(63.671,175.794,4.206)
			\Line[dash,dashsize=10](321,-212)(304,-148)
			\Line[dash,dashsize=10](449,-212)(464,-148)
			\Vertex(384,-154){8}
			\Vertex(304,-212){8}
			\end{picture}
			&
			\begin{picture}(194,74) (287,-223)
			\SetWidth{1.0}
			\SetColor{Black}
			\Line(288,-214)(480,-214)
			\Arc[dash,dashsize=10,clock](384.5,-218.669)(63.671,175.794,4.206)
			\Line[dash,dashsize=10](321,-214)(304,-150)
			\Line[dash,dashsize=10](449,-214)(464,-150)
			\Vertex(384,-214){8}
			\Vertex(304,-214){8}
			\end{picture}
			\\
			{\Huge $(U^{(2)}_{6})_{abij}$}
			&
			{\Huge $(U^{(2)}_{7})_{abij}$}
			&
			{\Huge $(U^{(2)}_{8})_{abij}$}
			&
			{\Huge $(U^{(2)}_{9})_{abij}$}
			&
			{\Huge $(U^{(2)}_{10})_{abij}$}
			\\
			&
			&
			&
			&
			\\
			\begin{picture}(194,66) (287,-223)
			\SetWidth{1.0}
			\SetColor{Black}
			\Line(288,-222)(480,-222)
			\Arc[dash,dashsize=10,clock](384.5,-226.669)(63.671,175.794,4.206)
			\Line[dash,dashsize=10](321,-222)(304,-158)
			\Line[dash,dashsize=10](449,-222)(464,-158)
			\CBox(321,-197)(301,-177){Black}{Black}
			\end{picture}
			&
			\begin{picture}(194,74) (287,-223)
			\SetWidth{1.0}
			\SetColor{Black}
			\Line(288,-214)(480,-214)
			\Arc[dash,dashsize=10,clock](384.5,-218.669)(63.671,175.794,4.206)
			\Line[dash,dashsize=10](321,-214)(304,-150)
			\Line[dash,dashsize=10](449,-214)(464,-150)
			\Vertex(309,-171){8}
			\Vertex(304,-214){8}
			\end{picture}
			&
			\begin{picture}(162,110) (303,-191)
			\SetWidth{1.0}
			\SetColor{Black}
			\Line(304,-178)(464,-178)
			\Line[dash,dashsize=10](336,-82)(384,-178)
			\Line[dash,dashsize=10](384,-178)(432,-82)
			\CBox(337,-190)(313,-166){Black}{Black}
			\Vertex(357,-178){8}
			\Vertex(411,-178){8}
			\end{picture}
			&
			\begin{picture}(162,110) (303,-191)
			\SetWidth{1.0}
			\SetColor{Black}
			\Line(304,-178)(464,-178)
			\Line[dash,dashsize=10](336,-82)(384,-178)
			\Line[dash,dashsize=10](384,-178)(432,-82)
			\CBox(349,-190)(325,-166){Black}{Black}
			\CBox(444,-190)(420,-166){Black}{Black}
			\end{picture}
			&
			\begin{picture}(162,106) (303,-191)
			\SetWidth{1.0}
			\SetColor{Black}
			\Line(304,-182)(464,-182)
			\Line[dash,dashsize=10](336,-86)(384,-182)
			\Line[dash,dashsize=10](384,-182)(432,-86)
			\Vertex(336,-182){8}
			\Vertex(432,-182){8}
			\CBox(365.314,-134.314)(342.686,-111.686){Black}{Black}
			\end{picture}
			\\
			{\Huge $(U^{(2)}_{11})_{abij}$}
			&
			{\Huge $(U^{(2)}_{12})_{abij}$}
			&
			{\Huge $(U^{(2)}_{13})_{abij}$}
			&
			{\Huge $(U^{(2)}_{14})_{abij}$}
			&
			{\Huge $(U^{(2)}_{15})_{abij}$}
			\\
			&
			&
			&
			&
			\\
			\begin{picture}(162,109) (303,-191)
			\SetWidth{1.0}
			\SetColor{Black}
			\Line(304,-179)(464,-179)
			\Line[dash,dashsize=10](336,-83)(384,-179)
			\Line[dash,dashsize=10](384,-179)(432,-83)
			\CBox(365.314,-131.314)(342.686,-108.686){Black}{Black}
			\CBox(347.314,-190.314)(324.686,-167.686){Black}{Black}
			\end{picture}
			&
			\begin{picture}(162,98) (303,-191)
			\SetWidth{1.0}
			\SetColor{Black}
			\Line(304,-190)(464,-190)
			\Line[dash,dashsize=10](336,-94)(384,-190)
			\Line[dash,dashsize=10](384,-190)(432,-94)
			\CBox(365.314,-142.314)(342.686,-119.686){Black}{Black}
			\CBox(425.314,-142.314)(402.686,-119.686){Black}{Black}
			\end{picture}
			&
			\begin{picture}(162,110) (303,-179)
			\SetWidth{1.0}
			\SetColor{Black}
			\Line(304,-178)(464,-178)
			\Arc[dash,dashsize=7.8](384,-146)(32.249,120,480)
			\Line[dash,dashsize=9.4](304,-82)(464,-82)
			\CBox(396,-128)(372,-104){Black}{Black}
			\CBox(396,-94)(372,-70){Black}{Black}
			\end{picture}
			&
			\begin{picture}(162,121) (303,-179)
			\SetWidth{1.0}
			\SetColor{Black}
			\Line(304,-167)(464,-167)
			\Arc(384,-113)(25.456,135,495)
			\CBox(394,-148)(374,-128){Black}{Black}
			\CBox(394,-178)(374,-158){Black}{Black}
			\Line[dash,dashsize=10](384,-87)(336,-55)
			\Line[dash,dashsize=10](384,-87)(432,-55)
			\end{picture}
			&
			\begin{picture}(162,153) (303,-207)
			\SetWidth{1.0}
			\SetColor{Black}
			\Line(304,-135)(464,-135)
			\Line[dash,dashsize=10](384,-135)(352,-55)
			\Line[dash,dashsize=10](384,-135)(416,-55)
			\Arc[dash,dashsize=7.4](384,-183)(22.627,135,495)
			\CBox(395.314,-174.314)(372.686,-151.686){Black}{Black}
			\Vertex(336,-135){8}
			\Vertex(432,-135){8}
			\end{picture}
			\\
			{\Huge $(U^{(2)}_{16})_{abij}$}
			&
			{\Huge $(U^{(2)}_{17})_{abij}$}
			&
			{\Huge $(U^{(2)}_{18})_{abij}$}
			&
			{\Huge $(U^{(2)}_{19})_{abij}$}
			&
			{\Huge $(U^{(2)}_{20})_{abij}$}
			\\
			&
			&
			&
			&
			\\
			\begin{picture}(162,153) (303,-207)
			\SetWidth{1.0}
			\SetColor{Black}
			\Line(304,-135)(464,-135)
			\Line[dash,dashsize=10](384,-135)(352,-55)
			\Line[dash,dashsize=10](384,-135)(416,-55)
			\CBox(395.314,-174.314)(372.686,-151.686){Black}{Black}
			\Vertex(336,-135){8}
			\Vertex(432,-135){8}
			\Arc(384,-183)(22.627,135,495)
			\end{picture}
			&
			\begin{picture}(162,114) (303,-175)
			\SetWidth{1.0}
			\SetColor{Black}
			\Line(304,-174)(464,-174)
			\Line[dash,dashsize=10](384,-174)(432,-62)
			\Line[dash,dashsize=10](384,-174)(336,-62)
			\CBox(361,-106)(337,-82){Black}{Black}
			\CBox(376,-142)(352,-118){Black}{Black}
			\end{picture}
			&
			\begin{picture}(178,126) (287,-175)
			\SetWidth{1.0}
			\SetColor{Black}
			\Line(288,-162)(464,-162)
			\Line[dash,dashsize=10](384,-162)(432,-50)
			\Line[dash,dashsize=10](384,-162)(336,-50)
			\CBox(332,-174)(308,-150){Black}{Black}
			\CBox(364,-174)(340,-150){Black}{Black}
			\end{picture}
			&
			\begin{picture}(162,137) (303,-152)
			\SetWidth{1.0}
			\SetColor{Black}
			\Line(304,-151)(464,-151)
			\Line[dash,dashsize=10](384,-151)(448,-39)
			\Line[dash,dashsize=10](384,-151)(320,-39)
			\CBox(360,-99)(336,-75){Black}{Black}
			\Arc[dash,dashsize=10](384,-39)(22.627,135,495)
			\CBox(396,-71)(372,-47){Black}{Black}
			\end{picture}
			&
			\begin{picture}(162,126) (303,-175)
			\SetWidth{1.0}
			\SetColor{Black}
			\Line(304,-162)(464,-162)
			\Line[dash,dashsize=10](416,-162)(464,-50)
			\Line[dash,dashsize=10](416,-162)(368,-50)
			\CBox(348,-174)(324,-150){Black}{Black}
			\CBox(348,-131)(324,-107){Black}{Black}
			\Arc[dash,dashsize=10](336,-98)(22.627,135,495)
			\end{picture}
			\\
			{\Huge $(U^{(2)}_{21})_{abij}$}
			&
			{\Huge $(U^{(2)}_{22})_{abij}$}
			&
			{\Huge $(U^{(2)}_{23})_{abij}$}
			&
			{\Huge $(U^{(2)}_{24})_{abij}$}
			&
			{\Huge $(U^{(2)}_{25})_{abij}$}
			\\
			&
			&
			&
			&
			\\
			\begin{picture}(162,90) (303,-211)
			\SetWidth{1.0}
			\SetColor{Black}
			\Line(304,-198)(464,-198)
			\CBox(364,-210)(340,-186){Black}{Black}
			\CBox(428,-210)(404,-186){Black}{Black}
			\Line[dash,dashsize=9.4](304,-134)(464,-134)
			\CBox(396,-146)(372,-122){Black}{Black}
			\end{picture}
			&
			\begin{picture}(162,90) (303,-211)
			\SetWidth{1.0}
			\SetColor{Black}
			\Line(304,-198)(464,-198)
			\CBox(364,-146)(340,-122){Black}{Black}
			\CBox(428,-146)(404,-122){Black}{Black}
			\Line[dash,dashsize=9.4](304,-134)(464,-134)
			\CBox(396,-210)(372,-186){Black}{Black}
			\end{picture}
			&
			\begin{picture}(162,115) (303,-184)
			\SetWidth{1.0}
			\SetColor{Black}
			\Line(304,-173)(464,-173)
			\CBox(394,-153)(374,-133){Black}{Black}
			\Line[dash,dashsize=9.4](304,-143)(464,-143)
			\CBox(394,-183)(374,-163){Black}{Black}
			\Arc[dash,dashsize=10](384,-93)(22.627,135,495)
			\CBox(394,-123)(374,-103){Black}{Black}
			\end{picture}
			&
			\begin{picture}(162,115) (303,-184)
			\SetWidth{1.0}
			\SetColor{Black}
			\Line(304,-173)(464,-173)
			\CBox(394,-153)(374,-133){Black}{Black}
			\Line[dash,dashsize=9.4](304,-143)(464,-143)
			\CBox(394,-183)(374,-163){Black}{Black}
			\CBox(394,-123)(374,-103){Black}{Black}
			\Arc(384,-93)(22.627,135,495)
			\end{picture}
			&
			\\
			{\Huge $(U^{(2)}_{26})_{abij}$}
			&
			{\Huge $(U^{(2)}_{27})_{abij}$}
			&
			{\Huge $(U^{(2)}_{28})_{abij}$}
			&
			{\Huge $(U^{(2)}_{29})_{abij}$}
			&
		\end{tabular*}
	}
	\caption{Two-loop tensor structures appearing in $\beta^{(2)}_{Y}$}
	\label{fig3}	
\end{table}
\clearpage

\begin{table}[t]
	\setlength{\extrarowheight}{1cm}
	\setlength{\tabcolsep}{24pt}
	\hspace*{-4.75cm}
	\centering
	\resizebox{6cm}{!}{
		\begin{tabular*}{20cm}{ccccc}
			\begin{picture}(162,114) (303,-223)
			\SetWidth{1.0}
			\SetColor{Black}
			\Line[dash,dashsize=9.4](304,-166)(464,-166)
			\Line[dash,dashsize=9.6](320,-110)(352,-166)
			\Line[dash,dashsize=9.6](448,-110)(416,-166)
			\Line[dash,dashsize=9.6](320,-222)(352,-166)
			\Line[dash,dashsize=9.6](448,-222)(416,-166)
			\Arc[dash,dashsize=10,clock](384,-165.015)(32.015,-178.237,-361.763)
			\Arc[dash,dashsize=10](384,-166.985)(32.015,178.237,361.763)
			\end{picture}
			&
			\begin{picture}(162,142) (303,-209)
			\SetWidth{1.0}
			\SetColor{Black}
			\Line[dash,dashsize=9.4](304,-138)(464,-138)
			\Line[dash,dashsize=9.6](344,-68)(424,-208)
			\Line[dash,dashsize=9.6](424,-68)(344,-208)
			\Arc[clock](384,-109.571)(27.571,150.513,29.487)
			\Arc(384,-82.429)(27.571,-150.513,-29.487)
			\end{picture}
			&
			\begin{picture}(161,142) (303,-209)
			\SetWidth{1.0}
			\SetColor{Black}
			\Line[dash,dashsize=10.6](304,-138)(464,-138)
			\Line[dash,dashsize=9.6](312,-68)(392,-208)
			\Line[dash,dashsize=9.6](392,-68)(312,-208)
			\Arc(416,-138)(21.932,114,474)
			\end{picture}
			&
			\begin{picture}(130,90) (335,-234)
			\SetWidth{1.0}
			\SetColor{Black}
			\Arc(400,-165)(42.521,139,499)
			\Arc[dash,dashsize=10,clock](400,-249)(70,143.13,36.87)
			\Line[dash,dashsize=10](376,-95)(368,-137)
			\Line[dash,dashsize=10](336,-109)(368,-137)
			\Line[dash,dashsize=10](424,-95)(432,-137)
			\Line[dash,dashsize=10](464,-109)(432,-137)
			\end{picture}
			&
			\begin{picture}(130,170) (335,-195)
			\SetWidth{1.0}
			\SetColor{Black}
			\Arc(400,-110)(42.521,139,499)
			\Line[dash,dashsize=10](336,-110)(464,-110)
			\Line[dash,dashsize=10.6](368,-26)(400,-67)
			\Line[dash,dashsize=10.6](432,-26)(400,-67)
			\Line[dash,dashsize=10.6](368,-194)(400,-153)
			\Line[dash,dashsize=10.6](432,-194)(400,-153)
			\end{picture}
			\\
			{\Huge $(V^{(2)}_{1})_{abij}$}
			&
			{\Huge $(V^{(2)}_{2})_{abij}$}
			&
			{\Huge $(V^{(2)}_{3})_{abij}$}
			&
			{\Huge $(V^{(2)}_{4})_{abij}$}
			&
			{\Huge $(V^{(2)}_{5})_{abij}$}
			\\
			&
			&
			&
			&
			\\
			\begin{picture}(162,142) (303,-209)
			\SetWidth{1.0}
			\SetColor{Black}
			\Line[dash,dashsize=9.4](304,-138)(464,-138)
			\Line[dash,dashsize=9.6](344,-68)(424,-208)
			\Line[dash,dashsize=9.6](424,-68)(344,-208)
			\CBox(371.662,-107.662)(348.338,-84.338){Black}{Black}
			\CBox(419.662,-107.662)(396.338,-84.338){Black}{Black}
			\end{picture}
			&
			\begin{picture}(161,142) (303,-209)
			\SetWidth{1.0}
			\SetColor{Black}
			\Line[dash,dashsize=10.6](304,-138)(464,-138)
			\Line[dash,dashsize=9.6](312,-68)(392,-208)
			\Line[dash,dashsize=9.6](392,-68)(312,-208)
			\CBox(411.662,-149.662)(388.338,-126.338){Black}{Black}
			\CBox(443.662,-149.662)(420.338,-126.338){Black}{Black}
			\end{picture}
			&
			\begin{picture}(161,142) (303,-209)
			\SetWidth{1.0}
			\SetColor{Black}
			\Line[dash,dashsize=10.6](304,-138)(464,-138)
			\Line[dash,dashsize=9.6](312,-68)(392,-208)
			\Line[dash,dashsize=9.6](392,-68)(312,-208)
			\CBox(443.662,-149.662)(420.338,-126.338){Black}{Black}
			\Arc[dash,dashsize=7](432,-88)(20.125,117,477)
			\CBox(443.662,-118.662)(420.338,-95.338){Black}{Black}
			\end{picture}
			&
			\begin{picture}(162,151) (303,-209)
			\SetWidth{1.0}
			\SetColor{Black}
			\Line[dash,dashsize=9.4](304,-129)(464,-129)
			\Line[dash,dashsize=9.6](344,-59)(384,-129)
			\Line[dash,dashsize=9.6](424,-59)(384,-129)
			\Arc[dash,dashsize=9](384,-153)(23.854,147,507)
			\Line[dash,dashsize=9.4](304,-199)(464,-199)
			\CBox(392.602,-183.602)(375.398,-166.398){Black}{Black}
			\CBox(392.602,-207.602)(375.398,-190.398){Black}{Black}
			\end{picture}
			&
			\begin{picture}(128,132) (335,-205)
			\SetWidth{1.0}
			\SetColor{Black}
			\Arc(400,-138)(42.521,139,499)
			\Line[dash,dashsize=10.6](368,-54)(400,-95)
			\Line[dash,dashsize=10.6](432,-54)(400,-95)
			\Line[dash,dashsize=10](336,-166)(368,-166)
			\Line[dash,dashsize=10](368,-166)(352,-194)
			\Line[dash,dashsize=10](432,-166)(464,-166)
			\Line[dash,dashsize=10](432,-166)(448,-194)
			\CBox(411.662,-193.662)(388.338,-170.338){Black}{Black}
			\end{picture}
			\\
			{\Huge $(V^{(2)}_{6})_{abij}$}
			&
			{\Huge $(V^{(2)}_{7})_{abij}$}
			&
			{\Huge $(V^{(2)}_{8})_{abij}$}
			&
			{\Huge $(V^{(2)}_{9})_{abij}$}
			&
			{\Huge $(V^{(2)}_{10})_{abij}$}
			\\
			&
			&
			&
			&
			\\
			\begin{picture}(128,140) (335,-195)
			\SetWidth{1.0}
			\SetColor{Black}
			\Arc(400,-140)(42.521,139,499)
			\Line[dash,dashsize=10.6](368,-56)(400,-97)
			\Line[dash,dashsize=10.6](432,-56)(400,-97)
			\Line[dash,dashsize=10](336,-168)(368,-168)
			\Line[dash,dashsize=10](368,-168)(352,-196)
			\Line[dash,dashsize=10](432,-168)(464,-168)
			\Line[dash,dashsize=10](432,-168)(448,-196)
			\Vertex(360,-126){8.602}
			\Vertex(440,-126){8.602}
			\end{picture}
			&
			\begin{picture}(146,127) (327,-197)
			\SetWidth{1.0}
			\SetColor{Black}
			\Arc(400,-153)(42.521,139,499)
			\Line[dash,dashsize=10](328,-111)(358,-153)
			\Line[dash,dashsize=10](358,-153)(328,-195)
			\Line[dash,dashsize=10](472,-111)(442,-153)
			\Line[dash,dashsize=10](442,-153)(472,-195)
			\Line[dash,dashsize=8.6](328,-83)(472,-83)
			\CBox(411.662,-124.662)(388.338,-101.338){Black}{Black}
			\CBox(411.662,-94.662)(388.338,-71.338){Black}{Black}
			\end{picture}
			&
			\begin{picture}(146,135) (327,-197)
			\SetWidth{1.0}
			\SetColor{Black}
			\Arc(400,-145)(42.521,139,499)
			\Line[dash,dashsize=10](328,-103)(358,-145)
			\Line[dash,dashsize=10](358,-145)(328,-187)
			\Line[dash,dashsize=10](472,-103)(442,-145)
			\Line[dash,dashsize=10](442,-145)(472,-187)
			\Line[dash,dashsize=8.6](328,-75)(472,-75)
			\CBox(411.662,-86.662)(388.338,-63.338){Black}{Black}
			\Vertex(400,-105){8.602}
			\Vertex(400,-187){8.602}
			\end{picture}
			&
			\begin{picture}(146,126) (327,-197)
			\SetWidth{1.0}
			\SetColor{Black}
			\Line[dash,dashsize=8.6](328,-84)(472,-84)
			\CBox(411.662,-95.662)(388.338,-72.338){Black}{Black}
			\Line[dash,dashsize=8.6](328,-196)(472,-196)
			\Line[dash,dashsize=8.6](328,-122)(472,-122)
			\CBox(411.662,-133.662)(388.338,-110.338){Black}{Black}
			\Arc(400,-176)(20.125,117,477)
			\CBox(411.662,-171.662)(388.338,-148.338){Black}{Black}
			\end{picture}
			&
			\\
			{\Huge $(V^{(2)}_{11})_{abij}$}
			&
			{\Huge $(V^{(2)}_{12})_{abij}$}
			&
			{\Huge $(V^{(2)}_{13})_{abij}$}
			&
			{\Huge $(V^{(2)}_{14})_{abij}$}
			&
		\end{tabular*}
	}
	\caption{Two-loop tensor structures appearing in $\beta^{(2)}_{h}$}
	\label{fig4}	
\end{table}

The computation of the two-loop Yukawa $\beta$-function is very straightforward and we simply quote the result, namely 
\begin{align}
\beta_Y^{(2)}=&8U^{(2)}_1+2(U^{(2)}_2+U^{(2)}_3)+\tfrac23(U^{(2)}_4+U^{(2)}_5)+8(U^{(2)}_6+U^{(2)}_7+U^{(2)}_8)\nn
&-16U^{(2)}_9+24U^{(2)}_{13}+4U^{(2)}_{14}+24U^{(2)}_{15}-16U^{(2)}_{16}-8(U^{(2)}_{17}+U^{(2)}_{18}+U^{(2)}_{19})\nn
&+2(U^{(2)}_{20}+U^{(2)}_{21})-\tfrac{40}{3}U^{(2)}_{22}-\tfrac43U^{(2)}_{23}-\tfrac83U^{(2)}_{24}-\tfrac23U^{(2)}_{25}\nn
&-32(U^{(2)}_{26}+U^{(2)}_{27})-8(U^{(2)}_{28}+U^{(2)}_{29}).
\label{valstwo}
\end{align}
Here and elsewhere we suppress a factor of $(8\pi)^{-1}$ for each loop order. The individual coefficients $c_{\alpha}^{(2)}$ defined in Eq.~\eqref{betdef} may then easily be read off, e.g. $c_1^{(2)}=8$ etc. 
The coefficients $c^{(2)}_{1}$, $c^{(2)}_{4}$, $c^{(2)}_{5}$ differ by factors from the corresponding results in Ref.~\cite{JJP} due to our slightly different choice of basis tensors here, as described above. Note further that with our choice of basis for
$U^{(2)}_6$--$U^{(2)}_{12}$, as explained in Appendix \ref{A1}, each non-vanishing term here corresponds to a single Feynman diagram.

There are $m_2=14$ tensor structures in the two-loop scalar $\beta$-function defined in Eq.~\eqref{betdef}, and they  are of the form depicted in Table~\ref{fig4} and defined precisely in Eq.~\eqref{scaltens}. Once again, in the case of $V_{8}^{(2)}$ there is also a graph with a fermion loop which is not depicted but whose contribution may be seen in Eq.~\eqref{scaltens}.The computation of the two-loop scalar $\beta$-function is again straightforward and it is given by
\begin{align}
\beta_h^{(2)}=&\tfrac{20}{3}V^{(2)}_1+30V^{(2)}_2+4V^{(2)}_3-360(V^{(2)}_4+V^{(2)}_5)-120V^{(2)}_6\nn
&-40V^{(2)}_7-8V^{(2)}_8
-120V^{(2)}_9+360(V^{(2)}_{10}-V^{(2)}_{11})+720(V^{(2)}_{13}+2V^{(2)}_{14}).
\label{scalvals}
\end{align}
Once again, the individual coefficients $d_{\alpha}^{(2)}$ defined in Eq.~\eqref{betdef} may easily be read off when required later.

We now turn to the construction of the $a$-function at lowest order. As discussed in Ref.~\cite{JJP}, we impose Eq.~\eqref{grad} in the form
\be
\frac{\pa A^{(5)}}{\pa Y_{abij}}=\beta^{(2)}_{abij},
\label{gradtwo}
\ee
where we define
\be
\frac{\pa }{\pa Y_{abij}}Y_{a'b'i'j'}=\tfrac14(\delta_{aa'}\delta_{bb'}+\delta_{ab'}\delta_{ba'})
(\delta_{ii'}\delta_{jj'}+\delta_{ij'}\delta_{ji'}).
\ee
The corresponding (lowest order) contribution to the metric $T_{IJ}$
is therefore effectively chosen to be the unit matrix in coupling space. The most general lowest-order $a$-function which may satisfy Eq.~\eqref{gradtwo}  is given by
\begin{align}
A^{(5)}=\sum_{\alpha=1}^9a^{(5)}_{\alpha}A^{(5)}_{\alpha}+\sum_{\alpha=13}^{29}a^{(5)}_{\alpha}A^{(5)}_{\alpha}
\label{Agen}
\end{align}
where the $A^{(5)}_{\alpha}$, $\alpha=1\ldots9$  are given in Eq.~\eqref{Adefs} and depicted in Table~\ref{fig0}. 
\begin{table}[t]
	\setlength{\extrarowheight}{1cm}
	\setlength{\tabcolsep}{24pt}
	\hspace*{-5.75cm}
	\centering
	\resizebox{6.7cm}{!}{
		\begin{tabular*}{20cm}{ccccc}
			\begin{picture}(162,162) (287,-207)
			\SetWidth{1.0}
			\SetColor{Black}
			\Arc(368,-126)(80,127,487)
			\Line[dash,dashsize=10](305,-78)(305,-175)
			\Line[dash,dashsize=9.4](305,-175)(433,-79)
			\Line[dash,dashsize=10](433,-79)(433,-173)
			\Line[dash,dashsize=8.2](433,-173)(305,-78)
			\end{picture}
			&
			\begin{picture}(162,162) (287,-207)
			\SetWidth{1.0}
			\SetColor{Black}
			\Arc(368,-126)(80,127,487)
			\Arc[dash,dashsize=10](493.891,-126.5)(148.892,147.728,212.272)
			\Arc[dash,dashsize=10](245.793,-125.728)(146.213,-33.299,32.578)
			\Arc[dash,dashsize=10,clock](368.272,-249.207)(146.213,123.299,57.422)
			\Arc[dash,dashsize=10,clock](365.917,-1.121)(148.893,-57.004,-121.554)
			\end{picture}
			&
			\begin{picture}(162,162) (287,-207)
			\SetWidth{1.0}
			\SetColor{Black}
			\Arc(368,-126)(80,127,487)
			\Arc[dash,dashsize=10](437.778,-126.5)(105.779,131.274,228.726)
			\Arc[dash,dashsize=10](302.985,-126.088)(103.019,-50.869,50.148)
			\Arc(369,-126)(37.443,146,506)
			\end{picture}
			&
			\begin{picture}(162,162) (287,-207)
			\SetWidth{1.0}
			\SetColor{Black}
			\Arc(368,-126)(80,127,487)
			\Arc[dash,dashsize=10](465.7,-126.5)(148.708,149.935,210.065)
			\Arc[dash,dashsize=10](159.567,-126.5)(192.439,-22.776,22.776)
			\Arc[dash,dashsize=10](540.898,-125.032)(156.898,153.081,207.715)
			\Arc[dash,dashsize=10](273.562,-125.598)(147.439,-29.41,28.609)
			\end{picture}
			&
			\begin{picture}(162,162) (191,-111)
			\SetWidth{1.0}
			\SetColor{Black}
			\Arc[dash,dashsize=8.4,clock](272.8,-30.4)(79.604,37.446,-90.576)
			\Arc[dash,dashsize=8.4,clock](273.109,-29.284)(80.723,-90.787,-206.711)
			\Arc[dash,dashsize=8.4,clock](271.587,-30.94)(81.019,152.077,33.689)
			\Arc(406.898,-31.816)(182.9,155.855,202.565)
			\Arc(106.024,-28.64)(150.995,-29.068,27.465)
			\Arc(465.342,-31.616)(177.344,155.474,203.736)
			\Arc(152.501,-29.459)(167.506,-26.042,25.252)
			\end{picture}
			\\
			{\Huge $A^{(5)}_{1}$}
			&
			{\Huge $A^{(5)}_{2}$}
			&
			{\Huge $A^{(5)}_{3}$}
			&
			{\Huge $A^{(5)}_{4}$}
			&
			{\Huge $A^{(5)}_{5}$}
			\\
			&
			&
			&
			&
			\\
			\begin{picture}(162,174) (287,-207)
			\SetWidth{1.0}
			\SetColor{Black}
			\Arc(368,-114)(80,127,487)
			\Arc[dash,dashsize=10,clock](215.28,-26.024)(152.928,-2.99,-56.808)
			\Arc[dash,dashsize=10](510.577,-31.856)(142.594,-179.138,-121.064)
			\Arc[dash,dashsize=10,clock](368,-263.025)(129.025,122.329,57.671)
			\CBox(380,-206)(356,-182){Black}{Black}
			\end{picture}
			&
			\begin{picture}(162,162) (287,-207)
			\SetWidth{1.0}
			\SetColor{Black}
			\Arc(368,-126)(80,127,487)
			\Arc[dash,dashsize=10,clock](215.28,-38.024)(152.928,-2.99,-56.808)
			\Arc[dash,dashsize=10](510.577,-43.856)(142.594,-179.138,-121.064)
			\Arc[dash,dashsize=10,clock](368,-275.025)(129.025,122.329,57.671)
			\CBox(379.402,-157.402)(356.598,-134.598){Black}{Black}
			\end{picture}
			&
			\begin{picture}(162,170) (287,-207)
			\SetWidth{1.0}
			\SetColor{Black}
			\Arc(368,-118)(80,127,487)
			\Arc[dash,dashsize=10,clock](215.28,-30.024)(152.928,-2.99,-56.808)
			\Arc[dash,dashsize=10](510.577,-35.856)(142.594,-179.138,-121.064)
			\Arc[dash,dashsize=10,clock](368,-267.025)(129.025,122.329,57.671)
			\Vertex(368,-138){7.81}
			\Vertex(368,-198){7.81}
			\end{picture}
			&
			\begin{picture}(162,162) (287,-207)
			\SetWidth{1.0}
			\SetColor{Black}
			\Arc(368,-126)(80,127,487)
			\Arc[dash,dashsize=10,clock](215.28,-38.024)(152.928,-2.99,-56.808)
			\Arc[dash,dashsize=10](510.577,-43.856)(142.594,-179.138,-121.064)
			\Arc[dash,dashsize=10,clock](368,-275.025)(129.025,122.329,57.671)
			\Vertex(304,-78){8}
			\Vertex(432,-78){8}
			\end{picture}
			&
			\\
			{\Huge $A^{(5)}_{6}$}
			&
			{\Huge $A^{(5)}_{7}$}
			&
			{\Huge $A^{(5)}_{8}$}
			&
			{\Huge $A^{(5)}_{9}$}
			&
		\end{tabular*}
	}
	\caption{Contributions to $A$ from Yukawa couplings}
	\label{fig0}	
\end{table}
The terms $A^{(5)}_{13}$--$A^{(5)}_{29}$ are defined by
\be
A^{(5)}_{\alpha}=(U^{(2)}_{\alpha})_{abij}Y_{abij},
\label{AU}
\ee

\noindent with $U^{(2)}_{\alpha}$ as defined in Eq.~\eqref{Yfour}; we choose not to display the corresponding diagrams which are of a very simple symmetric form. The reader will note that we have not defined structures corresponding to 
$A_{10}^{(5)}$-$A_{12}^{(5)}$. This is purely for notational convenience,  in order to ensure that $A^{(5)}_{\alpha}$ corresponds with $U^{(2)}_{\alpha}$ in Eq.~\eqref{AU}.
Then differentiating the terms in $A^{(5)}$  with respect to $Y_{abij}$  corresponds to removing each vertex in turn, leaving a structure which may be expressed in terms of one or possibly several of the ${U}^{(2)}_{\alpha}$. For instance:
 \begin{center}
 	\vspace*{-2cm}
 	\scalebox{0.45}
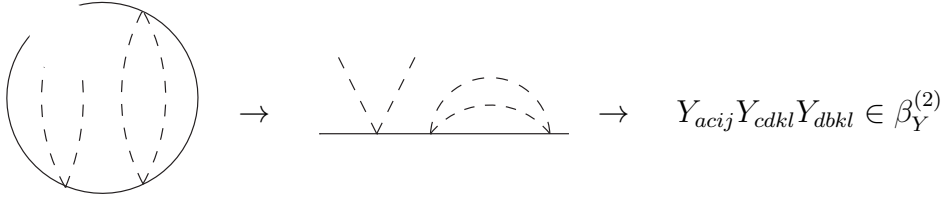
{\begin{picture}(162,232) (287,-147)
 		\SetWidth{1.0}
 		\SetColor{Black}
 		\Arc(368,-126)(80,127,487)
 		\Arc[dash,dashsize=10](465.75,-126.5)(148.751,149.945,210.055)
 		\Arc[dash,dashsize=10](159.5,-126.5)(192.501,-22.769,22.769)
 		\Arc[dash,dashsize=10](540.871,-125.032)(156.874,153.077,207.719)
 		\Arc[dash,dashsize=10](273.605,-125.599)(147.401,-29.419,28.617)
 		\SetColor{White}
 		\Vertex(337,-76){30}
 		\end{picture}
 	}$\;\;\;$ $\rightarrow$ $\;\;\;$
 	\scalebox{0.45}{\begin{picture}(201,85) (175,-166)
 		\SetWidth{1.0}
 		\SetColor{Black}
 		\Line(176,-175)(385,-175)
 		\Line[dash,dashsize=10](224,-175)(192,-111)
 		\Line[dash,dashsize=10](224,-175)(256,-111)
 		\Arc[dash,dashsize=10,clock](319,-178.096)(50.096,176.457,3.543)
 		\Arc[dash,dashsize=9,clock](319,-215.083)(64.083,141.282,38.718)
 		\end{picture}
 	}$\;\;\;$ $\rightarrow$ $\;\;\;$ $Y_{acij}Y_{cdkl}Y_{dbkl} \in \beta^{(2)}_{Y}$
 	\vspace*{1cm}
 \end{center}
 We then find immediately upon comparison of the coefficients of $U^{(2)}_{\alpha}$ in Eq.~\eqref{gradtwo} that
\be
a^{(5)}_{\alpha}=\tfrac14c^{(2)}_{\alpha},\quad \alpha=1,\ldots 5.
\label{aone}
\ee
There is no constraint on the two-loop coefficients $c^{(2)}_{1}$--$c^{(2)}_{5}$ in Eq.~\eqref{valstwo}. This is because the symmetries of the corresponding tensor structures appearing in Eq.~\eqref{Adefs} imply a one-to-one relation between $a$-function contributions and Yukawa $\beta$-function contributions. This may be seen in Table~\ref{fig0} where in each 
of the diagrams $A^{(5)}_{1}$-$A^{(5)}_{5}$, the removal of any vertex leads to the same $\beta$-function contribution. These $a$-function coefficients can thus be
tailored term-by-term to match any values for the $\beta$-function coefficients. The same is true of  $A^{(5)}_{13}$-$A^{(5)}_{29}$, and again we immediately find 
\begin{align}
a^{(5)}_{\alpha}&=\tfrac12c^{(2)}_{\alpha},\quad \alpha=13,\ldots25,\nn
a^{(5)}_{\alpha}&=c^{(2)}_{\alpha},\quad \alpha=26,\ldots29.
\label{athree}
\end{align}
However, the situation is different for 
$A^{(5)}_{6}$-$A^{(5)}_{9}$. There are four of these independent $a$-function structures, fewer than the seven $\beta$-function structures in our basis. Differentiation of each of these structures with respect to $Y_{abij}$ leads to two distinct
$\beta$-function structures, which in turn should be written in terms of our basis. Consequently we find a non-trivial set of equations relating $a^{(5)}_{6}$-$a^{(5)}_{9}$ with $c^{(2)}_6$-$c^{(2)}_{10}$ whose solution is
\be
a^{(5)}_6=a^{(5)}_8=c^{(2)}_6, \quad a^{(5)}_7=0,\quad a^{(5)}_9=c^{(2)}_8,
\label{atwo}
\ee
subject to the three consistency conditions
\be
c^{(2)}_6=c^{(2)}_7=-\tfrac12c^{(2)}_9,\quad c^{(2)}_8=c^{(2)}_7+\tfrac12c^{(2)}_{10},
\ee
which are indeed satisfied by the coefficients in Eq.~\eqref{valstwo}. In obtaining these equations, we imposed the vanishing of  two of the seven potential coefficients, namely $c^{(2)}_{11}$, $c^{(2)}_{12}$, which are manifestly zero for reasons explained in Appendix \ref{A1}. These consistency conditions correspond in the obvious way to relations among the simple poles in the Feynman diagrams corresponding to $U^{(2)}_6$-$U^{(2)}_{10}$. This is because, with our basis choice, the non-basis structures which could potentially have contributed to $c^{(2)}_6$-$c^{(2)}_{10}$ when written in terms of the basis, in fact correspond to vanishing Feynman diagram contributions.

Combining Eqs.~\eqref{aone}, \eqref{athree} and \eqref{atwo} with coefficients $c^{(2)}_{\alpha}$ read off from Eq.~\eqref{valstwo}, we find that the lowest-order contribution to the $a$-function is
\begin{align}
A^{(5)}=&2A^{(5)}_1+\tfrac12(A^{(5)}_2+A^{(5)}_3)+\tfrac16(A^{(5)}_4+A^{(5)}_5)+8(A^{(5)}_6+A^{(5)}_8+A^{(5)}_9)\nn
&+12A^{(5)}_{13}+2A^{(5)}_{14}+12A^{(5)}_{15}-8A^{(5)}_{16}-4(A^{(5)}_{17}+A^{(5)}_{18}+A^{(5)}_{19})\nn
&+A^{(5)}_{20}+A^{(5)}_{21}-\tfrac{20}{3}A^{(5)}_{22}-\tfrac23A^{(5)}_{23}-\tfrac43A^{(5)}_{24}-\tfrac13A^{(5)}_{25}\nn
&-32(A^{(5)}_{26}+A^{(5)}_{27})-8(A^{(5)}_{28}+A^{(5)}_{29}).
\end{align}
We cannot extend this leading order (i.e. five-loop) $a$-function in order to generate the two-loop scalar $\beta$-function in a similar way via Eq.~\eqref{grad}, as was pointed out in Ref.~\cite{JJP}. It is clear, on the analogy of the Yukawa $\beta$-function,  that in order to satisfy Eq.~\eqref{grad} with regard to the two-loop scalar $\beta$-function, the $a$-function must contain terms such as $(V^{(4)}_{\alpha})_{ijklmn}h_{ijklmn}$ (with $V^{(4)}_{\alpha}$ as defined in Eq.~\eqref{betdef}) which correspond to seven-loop diagrams.  We therefore postpone this discussion to the next section where we provide a full discussion of the seven-loop $a$-function.

\section{Higher-order results}
In this section we consider the seven-loop $a$-function, which will be determined via Eq.~\eqref{grad} by the two-loop scalar $\beta$-function and the four-loop Yukawa $\beta$-function. This hierarchy of loop orders for different couplings was first noticed in the four-dimensional context\cite{sann}. 
At this order we need to consider also next-to-leading order contributions on the right-hand side of Eq.\eqref{grad}.
Including all  relevant terms, we obtain (using a somewhat schematic notation)
\begin{eqnarray}
d_{h}A^{(7)} &=& dh\,T^{(5)}_{hh}\beta^{(2)}_{h} \label{dh}\\
d_{Y}A^{(7)} &=& dY\,T^{(5)}_{YY}\beta^{(2)}_{Y} + dY\,T^{(3)}_{YY}\beta^{(4)}_{Y} \label{dy}.
\label{gradseven}
\end{eqnarray}
Here $T^{(3)}_{YY}$  is the leading-order contribution to $T_{IJ}$ which as have mentioned before is effectively a unit tensor, and $T^{(5)}_{YY}$ is a potential higher order contribution described in detail later; we can easily see that no $T_{Yh}^{(5)}$ or $T_{hY}^{(5)}$ contributions are possible. We write
\be
A^{(7)}=A^{(7)}_{h}+A^{(7)}_{hY}+A^{(7)}_Y + a (\beta^{(2)}_{Y})_{abij}(\beta^{(2)}_{Y})_{abij}.
\label{Aseven}
\ee
Here $A^{(7)}_h$, $A^{(7)}_{hY}$ and $A^{(7)}_{Y}$ are the pure scalar, mixed scalar/Yukawa and pure Yukawa contributions to $A^{(7)}$,
respectively, while the last term represents the usual arbitrariness\cite{OsbJacnew} in the definition of $A$ satisying Eq.~\eqref{grad}. We remark that since the $\beta$-functions are renormalisation scheme-dependent beyond one loop, the $A$-function we construct will also be scheme-dependent. We analyse the scheme-dependence in more detail at the end of Appendix B. Here we simply note that in any particular scheme, at a critical point where the $\beta$-functions vanish, the dependence on $a$ in Eq.~\eqref{Aseven} disappears and the $A$-function is universal at the critical point (up to a numerical factor which we have fixed by requiring that the “metric” is the unit matrix at leading order). 

We can see how $\beta_h^{(2)}$ will determine $A^{(7)}_h$ and $A^{(7)}_{hY}$ through Eq.~\eqref{dh} while
$\beta^{(4)}_Y$ will determine $A^{(7)}_Y$ through Eq.~\eqref{dy}. Eq.~\eqref{dy} will also provide consistency checks on the mixed $a$-function terms $A^{(7)}_{hY}$. Starting with Eq.~\eqref{dh}, then, we expand
\be
 A^{(7)}_{h}=a^{(7)}_{h_1}A^{(7)}_{h_{1}},\quad A^{(7)}_{hY}=\sum\limits_{\alpha=2}^{14}a^{(7)}_{h_{\alpha}}A^{(7)}_{h_{\alpha}},
\label{adef}
 \ee
 as depicted in Table~\ref{fig1};
 while $A^{(7)}_Y$ will be defined later.
\begin{table}[t]
	\setlength{\extrarowheight}{1cm}
	\setlength{\tabcolsep}{24pt}
	\hspace*{-6.5cm}
	\centering
	\resizebox{7.5cm}{!}{
		\begin{tabular*}{20cm}{ccccc}
			\begin{picture}(162,162) (191,-111)
			\SetWidth{1.0}
			\SetColor{Black}
			\Arc[dash,dashsize=10](265.677,85.569)(101.765,-129.461,-43.903)
			\Arc[dash,dashsize=10](433.132,-117.254)(161.295,126.143,177.422)
			\Arc[dash,dashsize=10](113.213,-127.061)(159.701,6.133,57.082)
			\Arc[dash,dashsize=10,clock](272.8,-30.4)(79.604,37.446,-90.576)
			\Arc[dash,dashsize=10,clock](273.109,-29.284)(80.723,-90.787,-206.711)
			\Arc[dash,dashsize=10,clock](271.587,-30.94)(81.019,152.077,33.689)
			\Arc[dash,dashsize=10,clock](277.679,-186.2)(208.425,111.291,72.89)
			\Arc[dash,dashsize=10](363.385,26.038)(163.616,-172.965,-124.377)
			\Arc[dash,dashsize=10,clock](186.907,15.538)(151.101,-0.583,-56.183)
			\end{picture}
			&
			\begin{picture}(162,163) (191,-111)
			\SetWidth{1.0}
			\SetColor{Black}
			\Arc[dash,dashsize=10,clock](272.8,-29.4)(79.604,37.446,-90.576)
			\Arc[dash,dashsize=10,clock](273.109,-28.284)(80.723,-90.787,-206.711)
			\Arc[dash,dashsize=10,clock](271.587,-29.94)(81.019,152.077,33.689)
			\Arc[dash,dashsize=10](306.308,-29.5)(88.309,114.277,245.723)
			\Arc[dash,dashsize=10,clock](244.254,-28.691)(83.746,72.096,-73.528)
			\Arc[dash,dashsize=10](393.926,-29.778)(147.928,146.903,212.381)
			\Arc[dash,dashsize=10,clock](171.143,-29.5)(126.858,38.806,-38.806)
			\Arc(272,-29)(25.612,141,501)
			\Text(236,-30)[]{\Huge{\Black{$H_{1}$}}}
			\Text(312,-30)[]{\Huge{\Black{$H_{1}$}}}
			\end{picture}
			&
			\begin{picture}(162,162) (191,-111)
			\SetWidth{1.0}
			\SetColor{Black}
			\Arc[dash,dashsize=10,clock](272.8,-30.4)(79.604,37.446,-90.576)
			\Arc[dash,dashsize=10,clock](273.109,-29.284)(80.723,-90.787,-206.711)
			\Arc[dash,dashsize=10,clock](271.587,-30.94)(81.019,152.077,33.689)
			\Arc[dash,dashsize=10](313.5,-30.5)(91.501,118.386,241.614)
			\Arc[dash,dashsize=10,clock](238.309,-29.616)(85.692,68.295,-69.728)
			\Arc[dash,dashsize=10](443.614,-31.088)(191.617,154.965,204.319)
			\Arc[dash,dashsize=10,clock](137.364,-30.5)(154.637,30.938,-30.938)
			\Arc(292,-31)(21.471,118,478)
			\Text(276,5)[]{\Huge{\Black{$H_{2}$}}}
			\Text(276,-60)[]{\Huge{\Black{$H_{2}$}}}
			\end{picture}
			&
			\begin{picture}(162,162) (287,-207)
			\SetWidth{1.0}
			\SetColor{Black}
			\Arc(368,-126)(80,127,487)
			\Arc[dash,dashsize=10](301.579,-129.689)(67.474,2.284,76.789)
			\Arc[dash,dashsize=10](364.609,-76.876)(49.319,164.866,275.108)
			\Arc[dash,dashsize=10,clock](438.809,-130.465)(69.952,176.34,108.166)
			\Arc[dash,dashsize=10,clock](369.646,-77.704)(49.297,16.141,-89.589)
			\Arc[dash,dashsize=9.6](393.489,-215.871)(92.184,105.406,156.423)
			\Arc[dash,dashsize=9.6,clock](345.395,-210.674)(87.903,74.423,21.821)
			\Line[dash,dashsize=9](309,-181)(427,-181)
			\Text(305,-50)[]{\Huge{\Black{$H_{3}$}}}
			\Text(435,-50)[]{\Huge{\Black{$H_{3}$}}}
			\Text(305,-200)[]{\Huge{\Black{$H_{4}$}}}
			\Text(435,-200)[]{\Huge{\Black{$H_{4}$}}}
			\end{picture}
			&
			\begin{picture}(166,180) (285,-207)
			\SetWidth{1.0}
			\SetColor{Black}
			\Bezier[dash,dsize=8.8](369,-28)(441,-76)(292,-140)(369,-188)
			\Arc(368,-108)(80,127,487)
			\Bezier[dash,dsize=8.8](369,-28)(294,-76)(440,-140)(369,-188)
			\Line[dash,dashsize=9.4](289,-108)(448,-108)
			\Arc[dash,dashsize=10](368,-124.347)(81.653,168.451,371.549)
			\Text(370,-50)[]{\Huge{\Black{$H_{5}$}}}
			\Text(370,-165)[]{\Huge{\Black{$H_{5}$}}}
			\Text(270,-110)[]{\Huge{\Black{$H_{6}$}}}
			\Text(466,-110)[]{\Huge{\Black{$H_{6}$}}}
			\end{picture}
			\\
			{\Huge $A^{(7)}_{h_{1}}$}
			&
			{\Huge $A^{(7)}_{h_{2}}$}
			&
			{\Huge $A^{(7)}_{h_{3}}$}
			&
			{\Huge $A^{(7)}_{h_{4}}$}
			&
			{\Huge $A^{(7)}_{h_{5}}$}
			\\
			&
			&
			&
			&
			\\
			\begin{picture}(162,162) (287,-207)
			\SetWidth{1.0}
			\SetColor{Black}
			\Arc[dash,dashsize=10](368,-126)(79.906,134,494)
			\Arc[dash,dashsize=10](398.682,-126)(85.682,110.983,249.017)
			\Arc[dash,dashsize=10,clock](337.318,-126)(85.682,69.017,-69.017)
			\Arc[dash,dashsize=10](459.667,-126)(121.667,138.888,221.112)
			\Arc[dash,dashsize=10,clock](276.333,-126)(121.667,41.112,-41.112)
			\CBox(348,-136)(328,-116){Black}{Black}
			\CBox(408,-136)(388,-116){Black}{Black}
			\end{picture}
			&
			\begin{picture}(162,162) (287,-207)
			\SetWidth{1.0}
			\SetColor{Black}
			\Arc[dash,dashsize=10](368,-126)(79.906,134,494)
			\Arc[dash,dashsize=10](407,-126)(89,115.989,244.011)
			\Arc[dash,dashsize=10,clock](329,-126)(89,64.011,-64.011)
			\Arc[dash,dashsize=10](518,-126)(170,151.928,208.072)
			\Arc[dash,dashsize=10,clock](218,-126)(170,28.072,-28.072)
			\SetWidth{0.0}
			\CBox(397.314,-117.314)(374.686,-94.686){Black}{Black}
			\CBox(397.314,-157.314)(374.686,-134.686){Black}{Black}
			\end{picture}
			&
			\begin{picture}(162,162) (287,-207)
			\SetWidth{1.0}
			\SetColor{Black}
			\Arc[dash,dashsize=10](368,-126)(79.906,134,494)
			\Arc[dash,dashsize=10](391.333,-126)(83.333,106.26,253.74)
			\Arc[dash,dashsize=10,clock](344.667,-126)(83.333,73.74,-73.74)
			\Arc[dash,dashsize=10](441.929,-126)(108.929,132.741,227.259)
			\Arc[dash,dashsize=10,clock](294.071,-126)(108.929,47.259,-47.259)
			\SetWidth{0.0}
			\CBox(414.314,-137.314)(391.686,-114.686){Black}{Black}
			\SetWidth{1.0}
			\Arc[dash,dashsize=6](358,-126)(17.692,137,497)
			\SetWidth{0.0}
			\CBox(383.314,-137.314)(360.686,-114.686){Black}{Black}
			\end{picture}
			&
			\begin{picture}(162,172) (287,-197)
			\SetWidth{1.0}
			\SetColor{Black}
			\Arc[dash,dashsize=10](368,-116)(79.906,134,494)
			\Arc[dash,dashsize=10](443,-156)(85,151.928,208.072)
			\Arc[dash,dashsize=10,clock](293,-156)(85,28.072,-28.072)
			\Arc[dash,dashsize=10](379.667,-156)(41.667,106.26,253.74)
			\Arc[dash,dashsize=10,clock](356.333,-156)(41.667,73.74,-73.74)
			\Arc[dash,dashsize=8](425.75,-89)(63.75,154.942,205.058)
			\Arc[dash,dashsize=8,clock](310.25,-89)(63.75,25.058,-25.058)
			\SetWidth{0.0}
			\CBox(377.9,-78.899)(358.1,-59.101){Black}{Black}
			\CBox(377.9,-45.899)(358.1,-26.101){Black}{Black}
			\end{picture}
			&
			\begin{picture}(162,172) (287,-207)
			\SetWidth{1.0}
			\SetColor{Black}
			\Arc(368,-116)(80,127,487)
			\Arc[dash,dashsize=10](396.025,-77)(48.025,125.701,234.299)
			\Arc[dash,dashsize=10,clock](339.975,-77)(48.025,54.299,-54.299)
			\Arc[dash,dashsize=10](368.777,-182.364)(66.369,90.671,163.937)
			\Arc[dash,dashsize=10,clock](312.234,-108.151)(56.316,-8.012,-97.38)
			\Arc[dash,dashsize=10](423.766,-108.151)(56.316,-171.988,-82.62)
			\Arc[dash,dashsize=10,clock](365.427,-187.495)(71.541,87.939,19.172)
			\CBox(378,-206)(358,-186){Black}{Black}
			\Text(370,-20)[]{\Huge{\Black{$H_{7}$}}}
			\Text(285,-165)[]{\Huge{\Black{$H_{8}$}}}
			\Text(455,-165)[]{\Huge{\Black{$H_{8}$}}}
			\end{picture}
			\\
			{\Huge $A^{(7)}_{h_{6}}$}
			&
			{\Huge $A^{(7)}_{h_{7}}$}
			&
			{\Huge $A^{(7)}_{h_{8}}$}
			&
			{\Huge $A^{(7)}_{h_{9}}$}
			&
			{\Huge $A^{(7)}_{h_{10}}$}
			\\
			&
			&
			&
			&
			\\
			\begin{picture}(162,162) (287,-207)
			\SetWidth{1.0}
			\SetColor{Black}
			\Arc(368,-126)(80,127,487)
			\Arc[dash,dashsize=10](396.025,-87)(48.025,125.701,234.299)
			\Arc[dash,dashsize=10,clock](339.975,-87)(48.025,54.299,-54.299)
			\Arc[dash,dashsize=10](370.73,-194.927)(68.981,92.268,162.34)
			\Arc[dash,dashsize=10,clock](311.86,-117.66)(56.756,-8.45,-96.942)
			\Arc[dash,dashsize=10](424.14,-117.66)(56.756,-171.55,-83.058)
			\Arc[dash,dashsize=10,clock](365.427,-197.495)(71.541,87.939,19.172)
			\Vertex(310,-71){8.062}
			\Vertex(426,-71){8.062}
			\Text(370,-30)[]{\Huge{\Black{$H_{9}$}}}
			\Text(290,-185)[]{\Huge{\Black{$H_{10}$}}}
			\Text(455,-185)[]{\Huge{\Black{$H_{10}$}}}
			\end{picture}
			&
			\begin{picture}(162,172) (287,-197)
			\SetWidth{1.0}
			\SetColor{Black}
			\Arc(368,-116)(80,127,487)
			\Arc[dash,dashsize=7](383,-96)(25,126.87,233.13)
			\Arc[dash,dashsize=7,clock](353,-96)(25,53.13,-53.13)
			\Arc[dash,dashsize=10](370.73,-184.927)(68.981,92.268,162.34)
			\Arc[dash,dashsize=10,clock](311.86,-107.66)(56.756,-8.45,-96.942)
			\Arc[dash,dashsize=10](424.14,-107.66)(56.756,-171.55,-83.058)
			\Arc[dash,dashsize=10,clock](365.427,-187.495)(71.541,87.939,19.172)
			\SetWidth{0.0}
			\CBox(377.9,-85.899)(358.1,-66.101){Black}{Black}
			\CBox(377.9,-45.899)(358.1,-26.101){Black}{Black}
			\Text(290,-185)[]{\Huge{\Black{$H_{11}$}}}
			\Text(455,-185)[]{\Huge{\Black{$H_{11}$}}}
			\end{picture}
			&
			\begin{picture}(162,162) (287,-207)
			\SetWidth{1.0}
			\SetColor{Black}
			\Arc(368,-126)(80,127,487)
			\Arc[dash,dashsize=10,clock](408,-140.571)(42.571,159.984,20.016)
			\Arc[dash,dashsize=10](408,-111.429)(42.571,-159.984,-20.016)
			\Arc[dash,dashsize=8](341,-167.562)(49.562,56.991,123.009)
			\Arc[dash,dashsize=8,clock](341,-84.438)(49.562,-56.991,-123.009)
			\Arc[dash,dashsize=10](328,-140.571)(42.571,20.016,159.984)
			\Arc[dash,dashsize=10,clock](328,-111.429)(42.571,-20.016,-159.984)
			\CBox(330.9,-135.899)(311.1,-116.101){Black}{Black}
			\Vertex(321,-63){7.81}
			\Vertex(321,-189){7.81}
			\Text(270,-130)[]{\Huge{\Black{$H_{12}$}}}
			\Text(470,-130)[]{\Huge{\Black{$H_{12}$}}}
			\end{picture}
			&
			\begin{picture}(162,172) (287,-197)
			\SetWidth{1.0}
			\SetColor{Black}
			\Arc(368,-116)(80,127,487)
			\Arc[dash,dashsize=10](396.025,-77)(48.025,125.701,234.299)
			\Arc[dash,dashsize=10,clock](339.975,-77)(48.025,54.299,-54.299)
			\Arc[dash,dashsize=9.2](366.7,-153.9)(37.922,88.035,186.207)
			\Arc[dash,dashsize=8.4,clock](317.725,-108.423)(50.843,-8.57,-77.187)
			\Arc[dash,dashsize=10](418.275,-108.423)(50.843,-171.43,-102.813)
			\Arc[dash,dashsize=10,clock](369.3,-153.9)(37.922,91.965,-6.207)
			\SetWidth{0.0}
			\CBox(345.9,-159.899)(326.1,-140.101){Black}{Black}
			\CBox(409.9,-159.899)(390.1,-140.101){Black}{Black}
			\CBox(377.9,-205.899)(358.1,-186.101){Black}{Black}
			\Text(370,-23)[]{\Huge{\Black{$H_{13}$}}}
			\end{picture}
			&
			\\
			{\Huge $A^{(7)}_{h_{11}}$}
			&
			{\Huge $A^{(7)}_{h_{12}}$}
			&
			{\Huge $A^{(7)}_{h_{13}}$}
			&
			{\Huge $A^{(7)}_{h_{14}}$}
			&
		\end{tabular*}
	}
	\caption{Contributions to $A^{(7)}_{h}$, $A^{(7)}_{hY}$ }
	\label{fig1}
\end{table} 
Notice that, as suggested at the end of Section 2, we have
\be
A_{h_{\alpha}}^{(7)}=(V^{(4)}_{\alpha})_{ijklmn}h_{ijklmn},
\ee
with $V^{(4)}_{\alpha}$ as defined in Eq.~\eqref{betdef}.
The significance of the labels $H_1$-$H_{13}$ on the diagrams will be explained shortly.
  Eqs.~\eqref{betdef}, \eqref{dh}, \eqref{adef} then imply
 \begin{align}
 a^{(7)}_{h_1}&=\tfrac13\lambda d^{(2)}_1,\nn
a^{(7)}_{h_{\alpha}}&=\tfrac12\lambda d^{(2)}_{\alpha},\quad \alpha=2,3\quad \hbox{and} \quad \alpha=6,\ldots9, \nn
a^{(7)}_{h_{\alpha}}&=\lambda d^{(2)}_{\alpha},\quad \alpha=4,5 \quad \hbox{and} \quad \alpha=10\ldots14,
\label{Ahcoeff}
 \end{align}
so that $T^{(5)}_{hh}$ is the unit tensor up to a factor of $\lambda$ which will be determined shortly by the four-loop calculation. Reading off the coefficients $d_{\alpha}^{(2)}$ from Eq.~\eqref{scalvals}, and substituting Eq.~\eqref{Ahcoeff} into 
Eq.~\eqref{adef}, we find $A_h^{(7)}$ and  $A_{hY}^{(7)}$ in Eq.~\eqref{Aseven} are given by
\begin{align}
A_ h=&\tfrac{20}{9}\lambda A_{h_1}^{(7)},\nn
A_{hY}=&\lambda\Big[15A_{h_2}^{(7)}+2A_{h_3}^{(7)}-360(A_{h_4}^{(7)}+A_{h_5}^{(7)})-60A_{h_6}^{(7)}-20A_{h_7}^{(7)}-4A_{h_8}^{(7)}\nn
&-60A_{h_9}^{(7)}
+360(A_{h_{10}}^{(7)}-A_{h_{11}}^{(7)})+720(A_{h_{13}}^{(7)}+2A_{h_{14}}^{(7)})\Big].
\label{AhY}
\end{align}
In the main text we shall henceforth consider the ungauged case and omit $A_{h_6}^{(7)}$-$A_{h_{14}}^{(7)}$, in order to focus on our main purpose of providing evidence for the $a$-theorem in three dimensions. One piece of this evidence, presented already in Ref.~\cite{JJP}, is the consistency check mentioned earlier, arising from the fact that  $A^{(7)}_{hY}$ is determined by both $\beta^{(2)}_h$ and $\beta^{(4)}_Y$. However, if we regard the three-dimensional $a$-theorem as sufficiently established, we can use Eq.~\eqref{AhY} in conjunction with Eq.~\eqref{dy} to obtain a prediction for a hitherto-unknown part of the four-loop Yukawa $\beta$-function in the general gauged case. We shall postpone this to Appendix \ref{A3}.






Turning now to Eq.~\eqref{dy} in the ungauged case, we need to consider  in  Eq.~\eqref{Aseven} $A_{h_1}^{(7)}$-$A_{h_{14}}^{(7)}$
in $A_{hY}^{(7)}$, together with the pure Yukawa contributions, $A^{(7)}_Y$. We expand 
$A^{(7)}_Y$ as
\begin{equation}
A^{(7)}_Y = \sum\limits_{\alpha=1}^{52}a^{(7)}_{\alpha} A^{(7)}_{\alpha},
\label{A7Y}
\end{equation}
where the tensor structures $A^{(7)}_{\alpha}$ are depicted in Tables \ref{ATable1} and \ref{ATable2}. At this order we shall not give explicit expressions for the terms in the $a$-function since these are often quite unwieldy and may easily be reconstructed from the diagrams. In order to avoid having to specify all four-loop Yukawa $\beta$-function tensor structures explicitly, we have simply labelled every vertex in every $a$-function diagram in Tables \ref{fig1}, \ref{ATable1}, \ref{ATable2}. We now label each Yukawa $\beta$-function term according to the vertex which, when differentiated, yields that structure; denoting mixed gauge-Yukawa contributions from Tables \ref{fig1} by
$U^{(4)}_{H_{\alpha}}$ and pure Yukawa contributions from Tables \ref{ATable1}, \ref{ATable2} as $V^{(4)}_{\alpha}$. For instance, we have

\begin{center}
	\begin{table}[h]
		\setlength{\extrarowheight}{1cm}
		\setlength{\tabcolsep}{24pt}
		\hspace*{-10cm}
		\centering
		\resizebox{6cm}{!}{
\begin{tabular*}{20cm}{ccccc}
				\begin{picture}(265,213) (169,-141)
				\SetWidth{1.0}
				\SetColor{Black}
				\Arc(301,-39)(95.131,146,506)
				\Line[dash,dashsize=10](264,49)(359,-115)
				\Line[dash,dashsize=10](340,48)(245,-115)
				\Line[dash,dashsize=10](246,-115)(359,-115)
				\Line[dash,dashsize=10](264,49)(207,-52)
				\Line[dash,dashsize=10](207,-52)(396,-52)
				\Line[dash,dashsize=10](396,-52)(340,48)
				\Text(256,62)[]{\Huge{\Black{$1$}}}
				\Text(375,-133)[]{\Huge{\Black{$1$}}}
				\Text(237,-135)[]{\Huge{\Black{$1$}}}
				\Text(417,-52)[]{\Huge{\Black{$1$}}}
				\Text(184,-53)[]{\Huge{\Black{$1$}}}
				\Text(346,62)[]{\Huge{\Black{$1$}}}
				\end{picture}
				&
				\begin{picture}(231,238) (189,-132)
				\SetWidth{1.0}
				\SetColor{Black}
				\Arc(301,-14)(95.131,146,506)
				\Text(299,96)[]{\Huge{\Black{$2$}}}
				\Line[dash,dashsize=10](301,81)(220,-62)
				\Line[dash,dashsize=10](220,-62)(383,-62)
				\Line[dash,dashsize=10](383,-62)(301,81)
				\Line[dash,dashsize=10](220,35)(383,35)
				\Line[dash,dashsize=10](301,-108)(220,35)
				\Line[dash,dashsize=10](383,35)(301,-108)
				\Text(403,37)[]{\Huge{\Black{$2$}}}
				\Text(401,-76)[]{\Huge{\Black{$2$}}}
				\Text(204,41)[]{\Huge{\Black{$2$}}}
				\Text(204,-77)[]{\Huge{\Black{$2$}}}
				\Text(302,-126)[]{\Huge{\Black{$2$}}}
				\end{picture}
				&
				\begin{picture}(207,237) (196,-132)
				\SetWidth{1.0}
				\SetColor{Black}
				\Arc(301,-15)(95.131,146,506)
				\Text(299,95)[]{\Huge{\Black{$3$}}}
				\Arc[dash,dashsize=10](632.346,-15)(346.346,164.252,195.748)
				\Arc[dash,dashsize=10,clock](47.618,-15)(268.382,20.502,-20.502)
				\Arc[dash,dashsize=10](636.973,-397.437)(520.174,120.88,141.834)
				\Arc[dash,dashsize=10](85.866,228.62)(336.147,-64.987,-32.3)
				\Arc[dash,dashsize=10](53.141,-301.675)(390.064,35.169,63.366)
				\Arc[dash,dashsize=10,clock](579.031,309.036)(438.048,-118.205,-143.26)
				\Text(386,-91)[]{\Huge{\Black{$3$}}}
				\Text(212,-92)[]{\Huge{\Black{$3$}}}
				\Text(383,61)[]{\Huge{\Black{$3$}}}
				\Text(211,55)[]{\Huge{\Black{$3$}}}
				\Text(301,-126)[]{\Huge{\Black{$3$}}}
				\end{picture}
				&
				\begin{picture}(212,238) (196,-131)
				\SetWidth{1.0}
				\SetColor{Black}
				\Arc[dash,dashsize=10](301,-14)(95.425,128,488)
				\Arc(632.346,-14)(346.346,164.252,195.748)
				\Arc[clock](47.618,-14)(268.382,20.502,-20.502)
				\Arc[clock](579.031,310.036)(438.048,-118.205,-143.26)
				\Arc(53.141,-300.675)(390.064,35.169,63.366)
				\Arc(636.973,-396.437)(520.174,120.88,141.834)
				\Arc[clock](85.866,229.62)(336.147,-32.3,-64.987)
				\Text(299,97)[]{\Huge{\Black{$4$}}}
				\Text(299,-125)[]{\Huge{\Black{$4$}}}
				\Text(391,-87)[]{\Huge{\Black{$4$}}}
				\Text(211,-91)[]{\Huge{\Black{$4$}}}
				\Text(381,62)[]{\Huge{\Black{$4$}}}
				\Text(215,59)[]{\Huge{\Black{$4$}}}
				\end{picture}
				&
				\begin{picture}(260,211) (167,-156)
				\SetWidth{1.0}
				\SetColor{Black}
				\Arc(299,-41)(95.268,129,489)
				\Text(182,-31)[]{\Huge{\Black{$X$}}}
				\Text(410,-33)[]{\Huge{\Black{$X$}}}
				\Text(218,39)[]{\Huge{\Black{$X$}}}
				\Text(333,-148)[]{\Huge{\Black{$X$}}}
				\Text(267,-150)[]{\Huge{\Black{$X$}}}
				\Text(379,39)[]{\Huge{\Black{$X$}}}
				\Arc[dash,dashsize=9.6,clock](299,-175.133)(205.133,112.047,67.953)
				\Arc[dash,dashsize=9.6](299,205.133)(205.133,-112.047,-67.953)
				\Arc[dash,dashsize=10](382.861,15.268)(184.749,-165.496,-127.654)
				\Arc[dash,dashsize=10,clock](130.099,-151.555)(141.403,58.491,8.358)
				\Arc[dash,dashsize=10,clock](467.901,-151.555)(141.403,171.642,121.509)
				\Arc[dash,dashsize=10](215.139,15.268)(184.749,-52.346,-14.504)
				\end{picture}
				\\
\end{tabular*}
		}
		\caption{Contributions to $A^{(7)}$ - terms 1 to 30}
		\label{ATable1}
	\end{table}
\end{center}
\begin{center}
	\begin{table}[h]
		\setlength{\extrarowheight}{1cm}
		\setlength{\tabcolsep}{24pt}
		\hspace*{-10cm}
		\centering
		\resizebox{6.5cm}{!}{
			\begin{tabular*}{20cm}{ccccc}
				\begin{picture}(257,201) (171,-147)
				\SetWidth{1.0}
				\SetColor{Black}
				\Arc(299,-51)(95.268,129,489)
				\Text(240,44)[]{\Huge{\Black{$63$}}}
				\Text(186,-50)[]{\Huge{\Black{$X$}}}
				\Text(358,44)[]{\Huge{\Black{$63$}}}
				\Text(411,-50)[]{\Huge{\Black{$X$}}}
				\Text(394,-112)[]{\Huge{\Black{$64$}}}
				\Text(205,-113)[]{\Huge{\Black{$64$}}}
				\Line[dash,dashsize=10](245,27)(394,-51)
				\Line[dash,dashsize=10](245,27)(353,27)
				\Line[dash,dashsize=10](204,-51)(394,-51)
				\Line[dash,dashsize=10](353,27)(204,-51)
				\Arc[dash,dashsize=10,clock](299,-251.524)(174.524,118.397,61.603)
				\Arc[dash,dashsize=10](299,55.524)(174.524,-118.397,-61.603)
				\end{picture}
				&
				\begin{picture}(257,201) (171,-147)
				\SetWidth{1.0}
				\SetColor{Black}
				\Arc(299,-51)(95.268,129,489)
				\Text(240,44)[]{\Huge{\Black{$65$}}}
				\Text(186,-50)[]{\Huge{\Black{$X$}}}
				\Text(358,44)[]{\Huge{\Black{$65$}}}
				\Text(411,-50)[]{\Huge{\Black{$X$}}}
				\Text(394,-112)[]{\Huge{\Black{$66$}}}
				\Text(205,-113)[]{\Huge{\Black{$66$}}}
				\Arc[dash,dashsize=10,clock](299,-251.524)(174.524,118.397,61.603)
				\Arc[dash,dashsize=10](299,55.524)(174.524,-118.397,-61.603)
				\Arc[dash,dashsize=10,clock](355.947,-159.944)(186.968,144.36,90.903)
				\Arc[dash,dashsize=10](151.893,229.852)(285.645,-79.489,-45.247)
				\Arc[dash,dashsize=10,clock](242.053,-159.944)(186.968,89.097,35.64)
				\Arc[dash,dashsize=10](446.107,229.852)(285.645,-134.753,-100.511)
				\end{picture}
				&
				\begin{picture}(245,235) (177,-132)
				\SetWidth{1.0}
				\SetColor{Black}
				\Arc(299,-17)(95.268,129,489)
				\Text(298,93)[]{\Huge{\Black{$67$}}}
				\Arc[dash,dashsize=10](509.9,-17)(230.9,155.977,204.023)
				\Arc[dash,dashsize=10,clock](88.1,-17)(230.9,24.023,-24.023)
				\Arc[dash,dashsize=9.6,clock](299,-377.05)(401.05,102.822,77.178)
				\Arc[dash,dashsize=9.6](299,405.05)(401.05,-102.822,-77.178)
				\Arc[dash,dashsize=9.6](299,343.05)(401.05,-102.822,-77.178)
				\Arc[dash,dashsize=9.6,clock](299,-439.05)(401.05,102.822,77.178)
				\Text(299,-126)[]{\Huge{\Black{$67$}}}
				\Text(192,17)[]{\Huge{\Black{$68$}}}
				\Text(404,-54)[]{\Huge{\Black{$68$}}}
				\Text(193,-54)[]{\Huge{\Black{$68$}}}
				\Text(405,19)[]{\Huge{\Black{$68$}}}
				\end{picture}
				&
				\begin{picture}(196,235) (202,-132)
				\SetWidth{1.0}
				\SetColor{Black}
				\Arc(299,-17)(95.268,129,489)
				\Text(298,93)[]{\Huge{\Black{$X$}}}
				\Arc[dash,dashsize=10](509.9,-17)(230.9,155.977,204.023)
				\Arc[dash,dashsize=10,clock](88.1,-17)(230.9,24.023,-24.023)
				\Text(299,-126)[]{\Huge{\Black{$X$}}}
				\Text(218,63)[]{\Huge{\Black{$69$}}}
				\Text(381,-99)[]{\Huge{\Black{$69$}}}
				\Text(217,-98)[]{\Huge{\Black{$69$}}}
				\Text(379,63)[]{\Huge{\Black{$69$}}}
				\Arc[dash,dashsize=10](396.533,-17)(175.533,156.141,203.859)
				\Arc[dash,dashsize=10,clock](75.467,-17)(175.533,23.859,-23.859)
				\Arc[dash,dashsize=10,clock](201.467,-17)(175.533,23.859,-23.859)
				\Arc[dash,dashsize=10](522.533,-17)(175.533,156.141,203.859)
				\end{picture}
				&
				\begin{picture}(220,192) (203,-156)
				\SetWidth{1.0}
				\SetColor{Black}
				\Arc(299,-60)(95.268,129,489)
				\Text(218,20)[]{\Huge{\Black{$70$}}}
				\Line[dash,dashsize=10](232,7)(232,-127)
				\Line[dash,dashsize=10](232,7)(366,-127)
				\Line[dash,dashsize=10](366,7)(232,-127)
				\Line[dash,dashsize=10](366,7)(366,-127)
				\Arc(366,-60)(22.672,131,491)
				\Text(220,-139)[]{\Huge{\Black{$70$}}}
				\Text(376,22)[]{\Huge{\Black{$X$}}}
				\Text(377,-138)[]{\Huge{\Black{$X$}}}
				\Text(406,-42)[]{\Huge{\Black{$71$}}}
				\Text(406,-84)[]{\Huge{\Black{$71$}}}
				\end{picture}
				\\
\end{tabular*}
		}
		\caption{Contributions to $A^{(7)}$ - terms 31 to 52}
		\label{ATable2}
	\end{table}
\end{center}
\begin{center}
	\begin{table}[h]
		\setlength{\extrarowheight}{1cm}
		\setlength{\tabcolsep}{24pt}
		\hspace*{-8cm}
		\centering
		\resizebox{7cm}{!}{
			\begin{tabular*}{20cm}{ccccc}
				\begin{picture}(192,206) (203,-156)
				\SetWidth{2.0}
				\SetColor{Black}
				\Line(216,37)(246,7)
				\Line(246,37)(217,7)
				\SetWidth{1.0}
				\CTri(209.373,-110)(232,-87.373)(254.627,-110){Black}{Black}\CTri(209.373,-110)(232,-132.627)(254.627,-110){Black}{Black}
				\Text(231,-150)[]{\Huge{\Black{$(2)$}}}
				\Arc(299,-46)(94.895,119,479)
				\Line[dash,dashsize=10](232,21)(232,-113)
				\Line[dash,dashsize=10](232,21)(366,-113)
				\Line[dash,dashsize=10](366,21)(232,-113)
				\Line[dash,dashsize=10](366,21)(366,-113)
				\end{picture}
				&
				\begin{picture}(192,206) (203,-156)
				\SetWidth{2.0}
				\SetColor{Black}
				\Line(350,37)(380,7)
				\Line(380,37)(351,7)
				\SetWidth{1.0}
				\CTri(209.373,-110)(232,-87.373)(254.627,-110){Black}{Black}\CTri(209.373,-110)(232,-132.627)(254.627,-110){Black}{Black}
				\Text(231,-150)[]{\Huge{\Black{$(2)$}}}
				\Arc(299,-46)(94.895,119,479)
				\Line[dash,dashsize=10](232,21)(232,-113)
				\Line[dash,dashsize=10](232,21)(366,-113)
				\Line[dash,dashsize=10](366,21)(232,-113)
				\Line[dash,dashsize=10](366,21)(366,-113)
				\end{picture}
				&
				\begin{picture}(192,206) (203,-156)
				\SetWidth{2.0}
				\SetColor{Black}
				\Line(351,-97)(381,-127)
				\Line(381,-97)(352,-127)
				\SetWidth{1.0}
				\CTri(209.373,-110)(232,-87.373)(254.627,-110){Black}{Black}\CTri(209.373,-110)(232,-132.627)(254.627,-110){Black}{Black}
				\Text(231,-150)[]{\Huge{\Black{$(2)$}}}
				\Arc(299,-46)(94.895,119,479)
				\Line[dash,dashsize=10](232,21)(232,-113)
				\Line[dash,dashsize=10](232,21)(366,-113)
				\Line[dash,dashsize=10](366,21)(232,-113)
				\Line[dash,dashsize=10](366,21)(366,-113)
				\end{picture}
				&
				\begin{picture}(192,250) (203,-140)
				\SetWidth{2.0}
				\SetColor{Black}
				\Line(284,108)(314,78)
				\Line(314,108)(285,78)
				\SetWidth{1.0}
				\CTri(276.373,-94)(299,-71.373)(321.627,-94){Black}{Black}\CTri(276.373,-94)(299,-116.627)(321.627,-94){Black}{Black}
				\Text(298,-134)[]{\Huge{\Black{$(2)$}}}
				\Arc(299,-2)(94.895,119,479)
				\Line[dash,dashsize=10](299,92)(204,-2)
				\Line[dash,dashsize=10](299,-96)(204,-2)
				\Line[dash,dashsize=10](299,-96)(394,-2)
				\Line[dash,dashsize=10](299,92)(394,-2)
				\end{picture}
				&
				\begin{picture}(208,234) (187,-156)
				\SetWidth{2.0}
				\SetColor{Black}
				\Line(189,-2)(219,-32)
				\Line(219,-2)(190,-32)
				\SetWidth{1.0}
				\CTri(276.373,-110)(299,-87.373)(321.627,-110){Black}{Black}\CTri(276.373,-110)(299,-132.627)(321.627,-110){Black}{Black}
				\Text(298,-150)[]{\Huge{\Black{$(2)$}}}
				\Arc(299,-18)(94.895,119,479)
				\Line[dash,dashsize=10](299,76)(204,-18)
				\Line[dash,dashsize=10](299,-112)(204,-18)
				\Line[dash,dashsize=10](299,-112)(394,-18)
				\Line[dash,dashsize=10](299,76)(394,-18)
				\end{picture}
				\\
\end{tabular*}
		}
		\caption{Contributions to $T^{(5)}_{YY}$}
		\label{TTable}
	\end{table}
\end{center}
\clearpage

\be
\frac{\pa}{\pa Y_{abij}} A^{(7)}_{h_2}= 2(U^{(4)}_{H_1})_{abij},
\quad
\frac{\pa}{\pa Y_{abij}} A^{(7)}_{3}= 6(V^{(4)}_{3})_{abij},
\ee
where
\be
(U^{(4)}_{H_1})_{abij}=h_{ikmnpq}h_{jlmnpq}Y_{abkl},\quad
(V^{(4)}_{3})_{abij}=Y_{ackl}Y_{cdmn}Y_{deij}Y_{efkl}Y_{fbmn}.
\ee
An $X$ in Tables \ref{ATable1}, \ref{ATable2} corresponds to a structure which cannot occur in the $\beta$-function (for instance by virtue of being one-particle-reducible). The full four-loop Yukawa $\beta$-function accordingly takes the form
\be
\beta^{(4)}_{Y}=\sum_{\alpha=1}^{6}c_{H_{\alpha}}U^{(4)}_{H_{\alpha}}+ \sum_{\alpha=1}^{105}c_{\alpha}V^{(4)}_{\alpha},
\label{bydef}
\ee
writing separately the mixed scalar-Yukawa and pure Yukawa terms. Similarly, the associated next-to-leading-order metric takes the form
\begin{equation}
	T^{(5)}_{YY}=\sum\limits_{\alpha=1}^{18} t^{(5)}_{\alpha} (T_{\alpha}^{(5)})_{YY},
	\label{T5YY}
\end{equation}
where the tensor structures $T_{\alpha}^{(5)}$ are depicted in Table \ref{TTable} (which for convenience show $T_{\alpha}^{(5)}$ contracted with $dY$ and $\beta_Y$, denoted by a cross and a lozenge respectively). At this order $T_{IJ}$ is manifestly symmetric, as can easily be seen from the diagrams in Table \ref{TTable}; this was also the case at next-to-leading order in both four\cite{OsbJacnew,JacPoole} and six\cite{JP1} dimensions.

At two loops, in the non-gauge case, Eq.~\eqref{grad} imposed no constraints on the pure Yukawa contributions. By contrast, we shall find that Eq.~\eqref{dy}  imposes a large set of consistency conditions on the four-loop $\beta$-function coefficients, which we shall confirm by direct computation.   The full system of equations derived using (\ref{dy}) is highly non-trivial. Firstly we obtain
immediately from inserting $A^{(7)}_{hY}$ into Eq.~\eqref{dy} and comparing with Eq.~\eqref{bydef}:
\begin{align}
c^{(4)}_{H_1}&=\lambda d_2^{(2)}, &c^{(4)}_{H_2}&=\lambda d_3^{(2)},&
c^{(4)}_{H_3}&=2\lambda d_8^{(2)},\nn
c^{(4)}_{H_4}&=2\lambda d_8^{(2)},&
c^{(4)}_{H_5}&=2\lambda d_9^{(2)},& c^{(4)}_{H_6}&=2\lambda d_9^{(2)},
\label{cH}
\end{align}
relating four-loop Yukawa $\beta$-function coefficients to two-loop scalar $\beta$-function coefficients. These relations were checked already in Ref.~\cite{JJP}; they are satisfied provided $\lambda=\tfrac{1}{90}$.

The terms in $A^{(7)}_Y$ have been arranged such that substituting into (\ref{dy}) produces conditions in the following order:

\begin{itemize}
	\item Diagrams 1-6 simply relate the $a$-function coefficients $a^{(7)}_{1-6}$ to the $\beta^{(4)}_{Y}$ coefficients, and give no consistency conditions.
	\item Diagrams 7 and 8 relate tensor structures that appear in $\beta^{(4)}_{Y}$ to tensor structures that do not appear in $\beta^{(4)}_{Y}$, hence setting the corresponding $\beta$-function coefficients to zero.
	\item Diagrams 9-21 relate tensor structures that appear in $\beta^{(4)}_{Y}$ but not in any metric contributions, giving simple consistency conditions.
	\item Diagrams 22-47 relate tensor structures which appear both in $\beta^{(4)}_{Y}$ and in metric contributions, giving non-trivial consistency conditions.
	\item Diagrams 48-52, along with metric terms $t^{(5)}_{16-18}$ form a closed system of equations independent of the rest of the system.
\end{itemize}

\noindent With our choice of leading-order metric $T^{(3)}_{IJ}=\delta_{IJ}$, examples from the first four categories are:
\begin{itemize}
	\item Substituting $A^{(7)}_{1}$ gives $6a^{(7)}_{1} = 4 c^{(4)}_{1}$, so that $a^{(7)}_{1} = \tfrac23 c^{(4)}_{1}$ in a similar manner to the lowest-order calculation.
	\item Substituting $A^{(7)}_{7}$ gives $2a^{(7)}_{7}=2 c^{(4)}_{5}$ and $4a^{(7)}_{7}=0$, hence $c^{(4)}_{5}=0$.
	\item Substituting $A^{(7)}_{9}$ gives $2a^{(7)}_{7}=4c^{(4)}_{7}$, $2a^{(7)}_{7}=4 c^{(4)}_{8}$ and $2a^{(7)}_{7}=4 c^{(4)}_{9}$, hence $c^{(4)}_{7}=c^{(4)}_{8}=c^{(4)}_{9}$.
	\item Substituting $A^{(7)}_{25}$ and $A^{(7)}_{28}$ gives a set of nine equations:
	\begin{align}
		& a^{(7)}_{25}=4c^{(4)}_{45}, & & a^{(7)}_{25}=4 c^{(4)}_{46}, & & a^{(7)}_{25}=4c^{(4)}_{47}, \nn
		& a^{(7)}_{25}=8t^{(5)}_{5}, & & a^{(7)}_{25}=16t^{(5)}_{4}+4c^{(4)}_{48}, & & a^{(7)}_{25}=8t^{(5)}_{5}, \nn
		& & & & & \nn
		& 2a^{(7)}_{28}=4c^{(4)}_{58}, & & 2a^{(7)}_{28}=8t^{(5)}_{5}, & & 2a^{(7)}_{28}=16t^{(5)}_{4}+8t^{(5)}_{5}, \nn
	\end{align}
	leading to the consistency conditions $t^{(5)}_{4}=0$ and  $c^{(4)}_{45}=c^{(4)}_{46}=c^{(4)}_{47}=c^{(4)}_{48}=c^{(4)}_{58}$.
\end{itemize}

Deriving the full system of equations and eliminating the $a^{(7)}_{\alpha}$ and $t^{(5)}_{\alpha}$ coefficients therefore leaves a large set of consistency conditions on the $\beta^{(4)}_{Y}$ coefficients, which are given in full in Appendix \ref{A2}. We have computed these four-loop coefficients and checked that they satisfy all these conditions; again, the details are given in Appendix \ref{A2}. 
Finally, using the $\msbar$ values in Eqs.~(\ref{msbar}) and (\ref{AD}), the coefficients $a_i^{(7)}$  defined in Eq.~(\ref{A7Y}) may all be computed. Combining with the non-gauge parts of Eq.~\eqref{AhY} and subsituting into Eq.~\eqref{Aseven}, our final result for the $a$-function at next-to-leading order in the general scalar-fermion theory is
\begin{align}
A^{(7)}=&\tfrac{20}{9}A_{h_1}^{(7)}
+15A_{h_2}^{(7)}+2A_{h_3}^{(7)}
-360(A_{h_8}^{(7)}+A_{h_9}^{(7)})\nn
&-\tfrac43A^{(7)}_1+\tfrac{16}{3}A^{(7)}_2-\tfrac23A^{(7)}_3-\tfrac13A^{(7)}_4-\tfrac{7}{162}A^{(7)}_5+\tfrac{11}{162}A^{(7)}_6+(2\pi^2-16)A^{(7)}_9\nn
&+8A^{(7)}_{10}-2A^{(7)}_{11}+2\pi^2A^{(7)}_{12}+(\pi^2-8)A^{(7)}_{13}+\left(\tfrac83-16\pi^2\right)A^{(7)}_{14}+\tfrac{\pi^2}{2}A^{(7)}_{15}\nn&+4A^{(7)}_{16}+8A^{(7)}_{17}+\tfrac{\pi^2}{2}A^{(7)}_{19}+\pi^2A^{(7)}_{20}+2A^{(7)}_{21}+\pi^2A^{(7)}_{22}-\tfrac{16}{3}A^{(7)}_{23}-\tfrac{28}{3}A^{(7)}_{24}\nn&+2\pi^2A^{(7)}_{25}-\tfrac43A^{(7)}_{26}-\tfrac{56}{3}A^{(7)}_{27}
+\pi^2A^{(7)}_{28}-\tfrac{10}{9}A^{(7)}_{29}-\tfrac{28}{3}A^{(7)}_{30}
-\tfrac{16}{9}A^{(7)}_{31}-\tfrac89A^{(7)}_{32}\nn
&-\tfrac73A^{(7)}_{33}
-\tfrac{1}{54}A^{(7)}_{34}+\tfrac89A^{(7)}_{35}+\tfrac29A^{(7)}_{36}-\tfrac{14}{3}A^{(7)}_{37}-\tfrac29A^{(7)}_{38}+\tfrac{10}{27}A^{(7)}_{39}
-\tfrac29A^{(7)}_{40}
\nn&+\tfrac{11}{54}A^{(7)}_{41}+\tfrac{\pi^2}{6}(A^{(7)}_{42}+A^{(7)}_{43})-\tfrac83A^{(7)}_{44}+\pi^2A^{(7)}_{45}-\tfrac43A^{(7)}_{46}
+\tfrac19A^{(7)}_{47}+2\pi^2A^{(7)}_{48}\nn
&+\tfrac{\pi^2}{12}A^{(7)}_{49}+\tfrac{\pi^2}{4}A^{(7)}_{50}+\tfrac{\pi^2}{12}A^{(7)}_{51}
+\tfrac{\pi^2}{4}A^{(7)}_{52}.
\label{AY}
\end{align}
The next-to-leading-order metric coefficients in Eq.~(\ref{T5YY}) are likewise given by:
\begin{align}
t^{(5)}_{1}&=-\tfrac{28}{3}+3a, & t^{(5)}_{2}&=-\tfrac{40}{3}+3a, & t^{(5)}_{3}&=-\tfrac{16}{3}+3a, \nn
t^{(5)}_{4}&=0, & t^{(5)}_{5}&=\pi^2, &
t^{(5)}_{6}&=-\tfrac19+\tfrac14a, \nn
t^{(5)}_{7}+t^{(5)}_{8}&=-\tfrac23+\tfrac12a,& t^{(5)}_{9}&=-\tfrac{13}{3}+\tfrac34a, & t^{(5)}_{10}&=-\tfrac{14}{3}+\tfrac32a, \nn
t^{(5)}_{11}&=-\tfrac13+\tfrac34a,&
t^{(5)}_{12}&=-\tfrac{20}{3}+\tfrac12a, & t^{(5)}_{13}&=\tfrac{11}{9}+\tfrac34a,\nn t^{(5)}_{14}+t^{(5)}_{15}&=\tfrac12a,&t^{(5)}_{16}&=\tfrac{\pi^2}{4}, & t^{(5)}_{17}&=0, \nn
t^{(5)}_{18}&=\tfrac{\pi^2}{4}.&&&&&&&
\end{align} 
As in previous calculations in four\cite{OsbJacnew,JacPoole} and six{\cite{JP1} dimensions, we find that at each order the $a$-function is determined up to the  expected freedom  parametrised by $a$ in Eq.~\eqref{Aseven}. However, in the case of the metric, only the sums of $t^{(5)}_{7}$, $t^{(5)}_{8}$, and  $t^{(5)}_{14}$, $t^{(5)}_{15}$ are determined, leading to an additional arbitrariness. $t^{(5)}_{4}$ vanishes identically, whereas $t^{(5)}_{17}$ is proportional to $c^{(4)}_{103}-c^{(4)}_{105}$ and hence vanishes by Eq.~(\ref{cc6}).

\section{Conclusions} 
We have shown that Eq.~\eqref{grad} is valid at next-to-leading order for a general scalar-fermion theory. It seems possible that effectively we are constructing the $F$-function defined in Refs.\cite{jaff,klebb,kleba,klebc}. A proof was given in Ref.{\cite{klebb} that the $F$-function satisfies Eq.~\eqref{grad} in the neighbourhood of a conformal field theory at leading order; but our explicit results here provide further evidence beyond leading order (where, except as we have shown here for a gauge theory, the existence of an $a$-function with the properties of Eq.~\eqref{grad} is in any case trivial).  We have also shown that the two-loop scalar coupling $\beta$ function determines part of the next-to-leading order $a$-function which in turn gives a prediction for the scalar-coupling dependent sector of the four-loop Yukawa $\beta$-function, as displayed in Eq.~\eqref{pred}. We already checked the non-gauge part of this prediction in Ref.~\cite{JJP}; but we have now extended this calculation to include the gauge terms.
 
A natural next step would be to investigate the  case of $\Ncal=2$ supersymmetry, where partial results are available for the four-loop Yukawa $\beta$ function\cite{JL}. This would potentially extend the verification of Eq.~\eqref{grad} at next-to-leading order beyond the simple scalar-fermion case; and then assuming the equation applied more generally, it might be possible to deduce or at least strongly constrain the four-loop $\beta$ functions for a general gauge theory. One could further look for evidence for an all-orders form for the $a$-function in the $\Ncal=2$  supersymmetric case, such as was found in the four-dimensional case in  Refs.~\cite{AnselmiAM,BarnesJJ,KutasovXU,SusyA},\cite{OsbJacnew,JacPoole}.  The discussion in Ref.~\cite{morita} may provide some pointers in this direction.

\begin{table}[t]
	\setlength{\extrarowheight}{1cm}
	\setlength{\tabcolsep}{24pt}
	\hspace*{-5.75cm}
	\centering
	\resizebox{6.7cm}{!}{
		\begin{tabular*}{20cm}{ccccc}
			\begin{picture}(162,162) (287,-207)
			\SetWidth{1.0}
			\SetColor{Black}
			\Arc(368,-126)(80,127,487)
			\Arc(245.5,-50.75)(122.592,-65.16,2.221)
			\Arc(490.5,-50.75)(122.592,177.779,245.16)
			\Arc(503.236,-208.502)(211.413,129.768,167.293)
			\end{picture}
			&
			\begin{picture}(162,122) (287,-287)
			\SetWidth{1.0}
			\SetColor{Black}
			\Arc(328,-206)(39.962,122,482)
			\Arc(408,-206)(39.962,122,482)
			\Line(288,-206)(448,-206)
			\end{picture}
			&
			\begin{picture}(162,162) (287,-207)
			\SetWidth{1.0}
			\SetColor{Black}
			\Arc(245.5,-50.75)(122.592,-65.16,2.221)
			\Arc(368,-126)(80,127,487)
			\Arc(490.5,-50.75)(122.592,177.779,245.16)
			\Arc[clock](368,-311.531)(165.531,115.399,64.601)
			\end{picture}
			&
			\begin{picture}(162,162) (287,-207)
			\SetWidth{1.0}
			\SetColor{Black}
			\Arc(368,-126)(80,127,487)
			\Arc[clock](292.21,-151.793)(93.995,72.803,-35.219)
			\Arc(443.559,-150.665)(92.85,107.267,216.581)
			\Line(376,-73)(393,-73)
			\Line(393,-73)(393,-90)
			\Line(357,-67)(357,-84)
			\Line(357,-84)(341,-84)
			\end{picture}
			&
			\begin{picture}(162,177) (287,-207)
			\SetWidth{1.0}
			\SetColor{Black}
			\Arc(245.902,-35.996)(122.2,-65.282,2.343)
			\Arc(368,-111)(80,127,487)
			\Arc(490.098,-35.996)(122.2,177.657,245.282)
			\Arc[clock](368,-296.531)(165.531,115.399,64.601)
			\Line(344,-117)(359,-132)
			\Line(359,-132)(344,-147)
			\Line(344,-176)(359,-191)
			\Line(359,-191)(344,-206)
			\Line(368,-174)(368,-204)
			\end{picture}
			\\
			{\Huge $I_{4}$}
			&
			{\Huge $I_{22}$}
			&
			{\Huge $I_{4bbb}$}
			&
			{\Huge $I_{42bbc}$}
			&
			{\Huge $I_{42bb1de}$}
			\\
			&
			&
			&
			&
			\\
			\begin{picture}(176,178) (273,-207)
			\SetWidth{1.0}
			\SetColor{Black}
			\Arc(245.5,-34.75)(122.592,-65.16,2.221)
			\Arc(503.236,-192.502)(211.413,129.768,167.293)
			\Arc(368,-110)(80,127,487)
			\Arc(490.5,-34.75)(122.592,177.779,245.16)
			\Line(287,-57)(314,-77)
			\Line(293,-79)(274,-85)
			\Line(294,-79)(300,-98)
			\Line(354,-176)(369,-191)
			\Line(369,-191)(354,-206)
			\end{picture}
			&
			\begin{picture}(178,162) (271,-207)
			\SetWidth{1.0}
			\SetColor{Black}
			\Arc(368,-126)(80,127,487)
			\Line(368,-46)(368,-206)
			\Arc(428,-126)(100,126.87,233.13)
			\Arc[clock](308,-126)(100,53.13,-53.13)
			\Line(272,-126)(304,-126)
			\end{picture}
			&
			&
			&
			\\
			{\Huge $X$}
			&
			{\Huge $Y$}
			&
			&
			&
		\end{tabular*}
	}
	\caption{Frequently-occurring Feynman integrals}
	\label{momentum}	
\end{table}
   
The consistency conditions yield a set of relationships amongst the divergent contributions from three-dimensional Feynman integrals. As we have seen, for $A^{(7)}_{9}$-$A^{(7)}_{21}$ these conditions relate the simple poles in the diagrams obtained by deleting one of the vertices in a single $a$-function contribution. It was shown in Ref.~\cite{sax} that the Feynman diagrams obtained in the $\Ncal=2$ supersymmetric calculation could be expressed in terms of a relatively small set of integrals; and in fact this set can be further reduced using simple integration by parts. On the other hand in the current non-supersymmetric case additional basic integrals are required with double propagators. In Table \ref{momentum} we display the minimal  required set of integrals from Ref.~\cite{sax}, namely $I_4$, $I_{22}$, $I_{4bbb}$, $I_{42bbc}$, $I_{42bb1de}$ together with the most frequently occurring additional integrals, denoted $X$ and $Y$. Our consistency conditions from $A^{(7)}_{9}$-$A^{(7)}_{21}$  imply the extra relations 
\be
I_{42bbc}=I_4-\tfrac12I_{22}=-2X=-2Y,\quad I_{42bb1de}=-\tfrac12I_{4bbb},
\ee
so that everything can be expressed in terms of $I_4$, $I_{22}$ and $I_{4bbb}$. These relations cannot be obtained by any simple process of integration by parts and appear to be entirely new. They are also very easy to phrase diagrammatically at least for $A^{(7)}_{9}$-$A^{(7)}_{21}$. The diagrams of Table \ref{momentum} are in any case straightforward to compute using standard methods. However, it seems likely that the calculation for a general gauge theory would impose further relations among diagrams, possibly including those which are otherwise difficult to evaluate. It would be interesting to investigate whether there is any underlying topological explanation for these relations. 

{\Large{{\bf{Acknowledgements}}} }\hfil

We are very grateful to Tim Jones for useful conversations and for initiating our three-dimensional investigations.
This work was supported in part by the STFC under contract ST/G00062X/1, and CP was supported by an STFC studentship.

\appendix

\section{Tensor structures}
\label{A1}
In this appendix we give the exact expression for the various tensor structures which were described pictorially in the main text; and also explain in detail our choice of these tensor structures in view of relations among them resulting from gauge invariance as in Eq.~\eqref{gauge}. The contributions to the two-loop Yukawa $\beta$-function defined in Eq.~\eqref{betdef} were depicted in Table~\ref{fig3}. The explicit expressions for pure Yukawa terms are given by
\begin{align}
({U}_1^{(2)})_{abij}={}&\tfrac14[Y_{acil}Y_{cdjm}Y_{dblm}+Y_{aclm}Y_{cdjm}Y_{dbil}+Y_{acjl}Y_{cdim}Y_{dblm}
+Y_{aclm}Y_{cdim}Y_{dbjl}],\nn
({U}_2^{(2)})_{abij}={}&Y_{aclm}Y_{cdij}Y_{dblm},\nn
({U}_3^{(2)})_{abij}={}&Y_{cdik}Y_{abkl}Y_{cdlj},\nn
({U}_4^{(2)})_{abij}={}&\tfrac12[Y_{acij}Y_{cdlm}Y_{dblm}+Y_{adlm}Y_{dclm}Y_{cbij}],\nn
({U}_5^{(2)})_{abij}={}&\tfrac12[Y_{abik}Y_{cdkl}Y_{dclj}+Y_{cdil}Y_{dclk}Y_{abkj}],
\label{Ynone}
\end{align}
those with two gauge insertions are given by
\begin{align}
(U^{(2)}_6)_{abij}&=\tfrac14[Y_{acil}(E^{\psi2})_{cd}Y_{dblj}+(a\leftrightarrow b,\quad i\leftrightarrow j)],\nn
 (U^{(2)}_7)_{abij}&=\tfrac14[Y_{acil}E^{\psi}_{cd}E^{\phi}_{lm}Y_{dbmj}+(a\leftrightarrow b,\quad i\leftrightarrow j)],\nn
(U^{(2)}_8)_{abij}&=\tfrac14[E^{\psi}_{ac}Y_{cdil}Y_{delj}E^{\psi}_{be}+(a\leftrightarrow b,\quad i\leftrightarrow j)],\nn (U^{(2)}_9)_{abij}&=\tfrac14[E^{\psi}_{ac}Y_{cdil}E^{\phi}_{ml}Y_{dbmj}+(a\leftrightarrow b,\quad i\leftrightarrow j)],\nn
(U^{(2)}_{10})_{abij}&=\tfrac14[E^{\psi}_{ac}Y_{cdil}E^{\psi}_{de}Y_{eblj}+(a\leftrightarrow b,\quad i\leftrightarrow j)],\nn
(U^{(2)}_{11})_{abij}&=\tfrac14[(E^{\phi2})_{im}Y_{acml}Y_{cblj}+(a\leftrightarrow b,\quad i\leftrightarrow j)],\nn
(U^{(2)}_{12})_{abij}&=\tfrac14[E^{\psi}_{ac}E^{\phi}_{im}Y_{ceml}Y_{eblj}+(a\leftrightarrow b,\quad i\leftrightarrow j)],
\label{Ytwo}
\end{align}
and those with four or six by
\begin{align}
(U^{(2)}_{13})_{abij}=\tfrac12[(E^{\psi3})_{ac}Y_{cdij}E^{\psi}_{bd}+(a\leftrightarrow b)],&\quad(U^{(2)}_{14})_{abij}=\tfrac12[(E^{\psi2})_{ac}Y_{cdij}(E^{\psi2})_{db}+(a\leftrightarrow b)],\nn
(U^{(2)}_{15})_{abij}&=\tfrac14[(E^{\phi2})_{ik}E^{\psi}_{ac}Y_{cdkj}E^{\psi}_{bd}+(a\leftrightarrow b,\quad i\leftrightarrow j)],\nn
 (U^{(2)}_{16})_{abij}&=\tfrac14[(E^{\phi2})_{ik}(E^{\psi2})_{ac}Y_{cbkj}+(a\leftrightarrow b,\quad i\leftrightarrow j)],\nn
(U^{(2)}_{17})_{abij}=(E^{\phi2})_{ik}(E^{\phi2})_{jl}Y_{abkl}, &\quad (U^{(2)}_{18})_{abij}=(E^{\phi2})_{ij}(E^{\phi2})_{kl}Y_{abkl},\nn
(U^{(2)}_{19})_{abij}=(E^{\psi2})_{ab}(E^{\psi2})_{cd}Y_{cdij},&\quad (U^{(2)}_{20})_{abij}=\tr(E^{\phi2}) E^{\psi}_{ac}Y_{cdij}E^{\psi}_{bd},\nn
(U^{(2)}_{21})_{abij}=\tr(E^{\psi2}) E^{\psi}_{ac}Y_{cdij}E^{\psi}_{bd},&\quad (U^{(2)}_{22})_{abij}=\tfrac12[(E^{\phi4})_{ik}Y_{abkj}+(i\leftrightarrow j)],\nn
 (U^{(2)}_{23})_{abij}&=\tfrac12[(E^{\psi4})_{ac}Y_{cbij}+(a\leftrightarrow b)],\nn
(U^{(2)}_{24})_{abij}&=\tfrac12[\tr(E^{\phi2})+\tr(E^{\psi2})][(E^{\phi2})_{ik}Y_{abkj}+(i\leftrightarrow j)],\nn
(U^{(2)}_{25})_{abij}&=\tfrac12[\tr(E^{\phi2})+\tr(E^{\psi2})][(E^{\psi2})_{ac}Y_{cbij}+(a\leftrightarrow b)],\nn
(U^{(2)}_{26})_{abij}=(E^{\phi2})_{ij}(E^{\psi4})_{ab},&\quad
(U^{(2)}_{27})_{abij}=(E^{\phi4})_{ij}(E^{\psi2})_{ab},\nn
 (U^{(2)}_{28})_{abij}=\tr(E^{\phi2})(E^{\phi2})_{ij}(E^{\psi2})_{ab},&\quad
(U^{(2)}_{29})_{abij}=\tr(E^{\psi2})(E^{\phi2})_{ij}(E^{\psi2})_{ab}.
\label{Yfour}
\end{align}
Note that each tensor structure is symmetrised on $a$, $b$ and $i$, $j$ and has a ``weight'' of one. This is a different choice of basis convention from that adopted in Ref.~\cite{JJP}; but it means that differentiating a vertex in an $a$-function term gives a Yukawa $\beta$-function term with a factor of 1.  A word is in order regarding the choice of tensor structures in Eq.~\eqref{Ytwo}. There are in fact thirteen distinct tensor structures with two insertions of $E$;  however there are six identities amongst them resulting from application of Eq.~\eqref{gauge}, so that we can select seven of the structures as a basis. This will facilitate the construction of the $a$-function as we shall see shortly. Each of the thirteen structures corresponds to a two-loop Feynman diagram. However, eight of these manifestly give no $\beta$-function contribution, for the following reasons. Firstly, several of them are one-particle-reducible. For the remainder, the corresponding Feynman diagrams are logarithmically divergent so we can set the external momenta to zero. Structures corresponding to a Feyman diagram with only a single $\gamma$ matrix then manifestly give no contribution to the $\beta$-function by Lorentz invariance; and diagrams with an $E^{\phi}$ (and therefore a gauge vertex) on an external scalar line  give no contribution due to antisymmetry of the gauge propagator (resulting in a $\epsilon^{\mu\nu\rho}k_{\mu}k_{\nu}$ contribution). We have therefore selected the seven structures $U^{(2)}_{6}-U^{(2)}_{12}$ as a convenient basis since it includes all the diagrams (namely those corresponding to $U^{(2)}_{6}-U^{(2)}_{10}$) with potentially non-vanishing $\beta$-function contributions. Note that  in the abelian case which we are currently considering, some of them correspond to more than one Feynman diagram, with different orderings of the gauge matrices.

In the case of contributions with 4 and 6 insertions of $E^{\phi,\psi}$, we have not constructed a basis, since in this case there is a one-to-one equivalence between $\beta$-function and $a$-function contributions, and consequently the construction of the $a$-function is trivial. We have simply listed in Eq.~\eqref{Yfour} all the structures corresponding to non-vanishing contributions from Feynman diagrams. 

The contributions to the two-loop scalar $\beta$ function, as defined in Eq.~\eqref{betdef}, were depicted in Table~\ref{fig4} and are given explicitly by 
\begin{align}
 (V^{(2)}_1)_{ijklmn}=\tfrac{1}{6!}[h_{ijkpqr}h_{lmnpqr}+\hbox{perms}], \quad &
 (V^{(2)}_2)_{ijklmn}=\tfrac{1}{6!}[h_{ijklpq}Y_{abmp}Y_{abnq}+\hbox{perms}],\nn
 (V^{(2)}_3)_{ijklmn}=\tfrac{1}{6!}[h_{ijklmp}Y_{abpq}Y_{abnq}+\hbox{perms}],\quad&
(V^{(2)}_4)_{ijklmn}=\tfrac{1}{6!}[Y_{abij}Y_{bckl}Y_{cdmp}Y_{dapn}+\hbox{perms}], \nn
 (V^{(2)}_5)_{ijklmn}={}\tfrac{1}{6!}[Y_{abij}Y_{bcmp}Y_{cdkl}Y_{dapn}+\hbox{perms}],\quad&
 (V^{(2)}_6)_{ijklmn}={}\tfrac{1}{6!}[h_{ijklpq}(E^{\phi2})_{pm}(E^{\phi2})_{qn}+\hbox{perms}],\nn
(V^{(2)}_7)_{ijklmn}={}\tfrac{1}{6!}[h_{ijklmp}(E^{\phi4})_{pm}+\hbox{perms}],\quad&
(V^{(2)}_8)_{ijklmn}={}\tfrac{1}{6!}[h_{ijklmp}(E^{\phi2})_{pm}\{\tr(E^{\phi2})+\tr(E^{\psi2})\}\nn
&+\hbox{perms}],\nn
 (V^{(2)}_9)_{ijklmn}={}\tfrac{1}{6!}[h_{ijklpq}(E^{\phi2})_{qp}(E^{\phi2})_{mn}+\hbox{perms}],\quad&
(V^{(2)}_{10})_{ijklmn}=\tfrac{1}{6!}[Y_{abij}E^{\psi2}_{bc}Y_{cdkl}Y_{damn}+\hbox{perms}],\nn
 (V^{(2)}_{11})_{ijklmn}=\tfrac{1}{6!}[Y_{abij}E^{\psi}_{bc}Y_{cdkl}E^{\psi}_{de}Y_{eamn}+\hbox{perms}],\quad&
 (V^{(2)}_{12})_{ijklmn}=\tfrac{1}{6!}[E^{\phi2}_{ij}Y_{abkl}E^{\psi2}_{bc}Y_{camn}+\hbox{perms}],\nn
(V^{(2)}_{13})_{ijklmn}=\tfrac{1}{6!}[E^{\phi2}_{ij}Y_{abkl}E^{\psi}_{bc}Y_{cdmn}E^{\psi}_{da}+\hbox{perms}],\quad&
(V^{(2)}_{14})_{ijklmn}=\tfrac{1}{6!}[E^{\phi2}_{ij}E^{\phi2}_{kl}Y_{abmn}E^{\psi2}_{ba}+\hbox{perms}]
\label{scaltens}
\end{align}
 where ``+ perms'' completes the $6!$ permutations of the indices $\{ijklmn\}$. The choice of convention for the factors is similar to the Yukawa $\beta$-function terms.  Once again, and for similar reasons as in the case of Eq.~\eqref{Yfour}, we have not constructed a complete basis of independent tensor structures, but have simply listed in Eq.~\eqref{scaltens} all the structures corresponding to non-vanishing contributions from Feynman diagrams. 

The lowest-order (five-loop) $a$-function structures were depicted in Fig.~\ref{fig0} and given explicitly by
\begin{align}
A^{(5)}_1={}Y_{abij}Y_{bckl}Y_{cdik}Y_{dajl},&\quad
A^{(5)}_2={}Y_{abij}Y_{bckl}Y_{cdij}Y_{dakl},\nn
A^{(5)}_3={}Y_{abij}Y_{cdjk}Y_{abkl}Y_{cdli},&\quad
A^{(5)}_4={}Y_{acij}Y_{cbij}Y_{bdlm}Y_{dalm},\nn
A^{(5)}_5={}Y_{abik}Y_{bakj}Y_{cdil}Y_{dclj},&\quad
A^{(5)}_6={}Y_{abij}(E^{\psi2})_{bc}Y_{cdjk}Y_{daki},\nn
A^{(5)}_7={}Y_{abij}(E^{\phi2})_{jk}Y_{bckl}Y_{cali},&\quad
A^{(5)}_8={}Y_{abij}E^{\psi}_{bc}E^{\phi}_{jk}Y_{cdkl}Y_{dali},\nn
A^{(5)}_9={}&Y_{abij}E^{\psi}_{bc}Y_{cdjk}E^{\psi}_{de}Y_{eaki}.
\label{Adefs}
\end{align}
Note that we have for the non-gauge terms
\be
A^{(5)}_{\alpha}=(U_{\alpha}^{(4)})_{abij}Y_{abij},\quad \alpha=1,\ldots 5.
\ee

\section{Details of four-loop results}
\label{A2}
In this Appendix we list the full set of consistency conditions obtained by imposing Eq.~\eqref{grad} at next-to-leading  order, together with the detailed four-loop results for the Yukawa $\beta$-function coefficients. The simple consistency conditions 
obtained from $A^{(7)}_{9}$-$A^{(7)}_{21}$ are
\begin{align}
	c^{(4)}_{5}&=0, & c^{(4)}_{6}&=0, & c^{(4)}_{7}&=c^{(4)}_{8}=c^{(4)}_{9} ,\nn
	c^{(4)}_{10}&=c^{(4)}_{11}=c^{(4)}_{12}, & c^{(4)}_{13}&=2c^{(4)}_{14}, & c^{(4)}_{15}&=c^{(4)}_{16}=c^{(4)}_{17} ,\nn
	2c^{(4)}_{18}&=c^{(4)}_{19}, & 2c^{(4)}_{20}&=c^{(4)}_{21}, & 2c^{(4)}_{22}&=c^{(4)}_{23} ,\nn
	c^{(4)}_{24}&=2c^{(4)}_{25}, & 2c^{(4)}_{26}&=c^{(4)}_{27}, & c^{(4)}_{28}&=c^{(4)}_{29} ,\nn
	2c^{(4)}_{30}&=c^{(4)}_{31}, & 2c^{(4)}_{32}&=c^{(4)}_{33}, & c^{(4)}_{34}&=2c^{(4)}_{35} ,\nn
	c^{(4)}_{54}&=c^{(4)}_{56}, & c^{(4)}_{89}&=c^{(4)}_{90}, & &
	\label{cc1}
\end{align}
and those resulting from $A^{(7)}_{22}$-$A^{(7)}_{47}$ are 
\begin{align}
c^{(4)}_{40}-c^{(4)}_{39}&=c^{(4)}_{42}-c^{(4)}_{44}=c^{(4)}_{50}-c^{(4)}_{49}\nn
&=c^{(4)}_{52}-c^{(4)}_{57}=3(c^{(4)}_{70}-c^{(4)}_{72})=c^{(4)}_{87}-c^{(4)}_{86}, \nn
c^{(4)}_{40}-c^{(4)}_{41}&=c^{(4)}_{42}-c^{(4)}_{43}=c^{(4)}_{50}-c^{(4)}_{51}=c^{(4)}_{52}-c^{(4)}_{53}\nn
&=6(c^{(4)}_{70}-c^{(4)}_{72})-c^{(4)}_{61}+c^{(4)}_{62}=c^{(4)}_{87}-c^{(4)}_{88},\nn
c^{(4)}_{55}-4c^{(4)}_{67}=\tfrac12(c^{(4)}_{56}-2c^{(4)}_{68}), &\quad 3(c^{(4)}_{70}+c^{(4)}_{72})+c^{(4)}_{85}=c^{(4)}_{88}+12c^{(4)}_{97} ,\nn
c^{(4)}_{75}=12(c^{(4)}_{80}-c^{(4)}_{97}), & \quad c^{(4)}_{85}-c^{(4)}_{89}=2(2c^{(4)}_{94}-c^{(4)}_{95}),
\end{align}
and
\begin{align}
c^{(4)}_{52}-c^{(4)}_{55}-6c^{(4)}_{63}+12c^{(4)}_{65}+6c^{(4)}_{70}+c^{(4)}_{85}-c^{(4)}_{87}-12c^{(4)}_{97}&=0 ,\nn
3c^{(4)}_{59}+3c^{(4)}_{70}-6c^{(4)}_{77}-c^{(4)}_{87}+c^{(4)}_{89}-6c^{(4)}_{97}&=0 ,\nn
6c^{(4)}_{63}-3c^{(4)}_{70}+3c^{(4)}_{72}-12c^{(4)}_{77}-c^{(4)}_{85}-c^{(4)}_{88}+2c^{(4)}_{89}&=0 ,\nn
3c^{(4)}_{65}+c^{(4)}_{74}-3c^{(4)}_{77}&=0 ,\nn
4c^{(4)}_{67}-c^{(4)}_{68}+2c^{(4)}_{74}+6c^{(4)}_{80}+4c^{(4)}_{94}-2c^{(4)}_{95}-6c^{(4)}_{97}&=0,\nn
2c^{(4)}_{36}=c^{(4)}_{37}=2c^{(4)}_{38}=c^{(4)}_{45}=c^{(4)}_{46}=c^{(4)}_{47}=c^{(4)}_{48}=c^{(4)}_{58}=2c^{(4)}_{91}=c^{(4)}_{92}&=2c^{(4)}_{93}.
\label{cc5}
\end{align}
The condition resulting from $A^{(7)}_{48}$-$A^{(7)}_{52}$ is 
\begin{equation}
c^{(4)}_{98}=c^{(4)}_{99}=c^{(4)}_{100}=c^{(4)}_{101}=6c^{(4)}_{102}=4c^{(4)}_{103}=6c^{(4)}_{104}=4c^{(4)}_{105}.
\label{cc6}
\end{equation}
All anomalous dimension terms (except $c^{(4)}_{102}$ and $c^{(4)}_{104}$) have been eliminated from the above consistency conditions; the consistency conditions which do involve anomalous dimension coefficients may be expressed in the following form. 
\begin{align}
c^{(4)}_{60}&=c^{(4)}_{59}-2c^{(4)}_{63}+12c^{(4)}_{69}+2c^{(4)}_{70}-\tfrac16c^{(4)}_{86}-\tfrac16c^{(4)}_{87}+\tfrac13c^{(4)}_{90}+\tfrac43c^{(4)}_{94}-\tfrac23c^{(4)}_{95}-4c^{(4)}_{97} ,\nn
c^{(4)}_{64}&=-c^{(4)}_{63}+12c^{(4)}_{69}+c^{(4)}_{70}-\tfrac16c^{(4)}_{86}+\tfrac16c^{(4)}_{90}+\tfrac23c^{(4)}_{94}-\tfrac13c^{(4)}_{95}-2c^{(4)}_{97} ,\nn
c^{(4)}_{66}&=-\tfrac12c^{(4)}_{63}+6c^{(4)}_{69}+\tfrac12c^{(4)}_{70}+\tfrac{1}{6}c^{(4)}_{74}-\tfrac12c^{(4)}_{80}+\tfrac13c^{(4)}_{94}-\tfrac16c^{(4)}_{95}-\tfrac12c^{(4)}_{97} ,\nn
c^{(4)}_{71}&=12c^{(4)}_{82}-\tfrac16c^{(4)}_{86}+\tfrac16c^{(4)}_{90}+\tfrac23c^{(4)}_{94}-\tfrac13c^{(4)}_{95}-2c^{(4)}_{97} ,\nn
c^{(4)}_{73}&=12c^{(4)}_{82}-\tfrac16c^{(4)}_{87}+\tfrac16c^{(4)}_{90}+\tfrac23c^{(4)}_{94}-\tfrac13c^{(4)}_{95}-2c^{(4)}_{97} ,\nn
c^{(4)}_{76}&=-c^{(4)}_{63}+6c^{(4)}_{69}+c^{(4)}_{70}+c^{(4)}_{77}+\tfrac23c^{(4)}_{94}-\tfrac13c^{(4)}_{95}-2c^{(4)}_{97} ,\nn
c^{(4)}_{78}&=-\tfrac16c^{(4)}_{63}+2c^{(4)}_{69}+\tfrac16c^{(4)}_{70} ,\nn
c^{(4)}_{79}&=\tfrac16c^{(4)}_{63}-\tfrac16c^{(4)}_{70}+2c^{(4)}_{82} ,\nn
c^{(4)}_{81}&=\tfrac13c^{(4)}_{74}+6c^{(4)}_{82}+\tfrac23c^{(4)}_{94}-\tfrac13c^{(4)}_{95}-c^{(4)}_{97} ,\nn
c^{(4)}_{83}&=\tfrac16c^{(4)}_{92} ,\nn
c^{(4)}_{84}&=\tfrac16c^{(4)}_{92} ,\nn
c^{(4)}_{96}&=6c^{(4)}_{82}+\tfrac13c^{(4)}_{94}-\tfrac16c^{(4)}_{95}-c^{(4)}_{97}.
\label{ADpredict}
\end{align}
We see that in this form these conditions have the effect of predicting the values of all the anomalous dimension coefficients except two, namely $c^{(4)}_{69}$ and $c^{(4)}_{82}$. We need only calculate the two explicit Feynman integrals depicted in Table \ref{fig2} corresponding to $c^{(4)}_{69}$ and $c^{(4)}_{82}$ in order to obtain all 14 possible anomalous dimension coefficients. This is useful since the Feynman diagrams corresponding to anomalous dimensions are typically harder to evaluate than the others, being linearly or quadratically divergent.

\begin{center}
	\begin{table}[t]
		\setlength{\extrarowheight}{1cm}
		\setlength{\tabcolsep}{24pt}
		\hspace*{1.25cm}
		\centering
		\resizebox{8cm}{!}{
			\begin{tabular*}{20cm}{cc}
				\begin{picture}(182,122) (199,-195)
				\SetWidth{1.0}
				\SetColor{Black}
				\Line(200,-134)(380,-134)
				\Arc[dash,dashsize=10](290,-134)(29.833,140,500)
				\Arc[dash,dashsize=10](290,-134)(59.816,142,502)
				\end{picture}
				&
				\begin{picture}(172,120) (199,-196)
				\SetWidth{1.0}
				\SetColor{Black}
				\Line[dash,dashsize=10](200,-136)(380,-136)
				\Arc(290,-136)(30.414,134,494)
				\Arc(290,-136)(59.414,134,494)
				\end{picture}
				\\
				{\Huge $c^{(4)}_{69}$}
				&
				{\Huge $c^{(4)}_{82}$}
			\end{tabular*}
		}
		\caption{Feynman integrals for terms undetermined by \eqref{dy}}
		\label{fig2}
	\end{table}
\end{center}

The non-anomalous dimension terms can all be calculated in $\msbar$ via integration by parts, using master integrals computed in Ref.~\cite{sax}. The coefficients are:
\begin{align}
c^{(4)}_{1}&=-8, & c^{(4)}_{2}&=32, & c^{(4)}_{3}&=-4, & c^{(4)}_{4}&=-2, & c^{(4)}_{5}&=0, \nn
c^{(4)}_{6}&=0, & c^{(4)}_{7}&=4(\pi^2-8), & c^{(4)}_{8}&=4(\pi^2-8), & c^{(4)}_{9}&=4(\pi^2-8), & c^{(4)}_{10}&=16, \nn
c^{(4)}_{11}&=16, & c^{(4)}_{12}&=16, & c^{(4)}_{13}&=-8, & c^{(4)}_{14}&=-4, & c^{(4)}_{15}&=4\pi^2, \nn
c^{(4)}_{16}&=4\pi^2, & c^{(4)}_{17}&=4\pi^2, & c^{(4)}_{18}&=2(\pi^2-8), & c^{(4)}_{19}&=4(\pi^2-8), & c^{(4)}_{20}&=16\left(\tfrac{\pi^2}{3}-2\right), \nn
c^{(4)}_{21}&=32\left(\tfrac{\pi^2}{3}-2\right), & c^{(4)}_{22}&=\pi^2, & c^{(4)}_{23}&=2\pi^2, & c^{(4)}_{24}&=16, & c^{(4)}_{25}&=8, \nn
c^{(4)}_{26}&=32, & c^{(4)}_{27}&=64, & c^{(4)}_{28}&=0, & c^{(4)}_{29}&=0, & c^{(4)}_{30}&=\pi^2, \nn
c^{(4)}_{31}&=2\pi^2, & c^{(4)}_{32}&=2\pi^2, & c^{(4)}_{33}&=4\pi^2, & c^{(4)}_{34}&=8, & c^{(4)}_{35}&=4, \nn
c^{(4)}_{36}&=\pi^2, & c^{(4)}_{37}&=2\pi^2, & c^{(4)}_{38}&=\pi^2, & c^{(4)}_{39}&=8, & c^{(4)}_{40}&=16, \nn
c^{(4)}_{41}&=0, & c^{(4)}_{42}&=8, & c^{(4)}_{43}&=-8, & c^{(4)}_{44}&=0, & c^{(4)}_{45}&=2\pi^2, \nn
c^{(4)}_{46}&=2\pi^2, & c^{(4)}_{47}&=2\pi^2, & c^{(4)}_{48}&=2\pi^2, & c^{(4)}_{49}&=16, & c^{(4)}_{50}&=24\nn
c^{(4)}_{51}&=8, & c^{(4)}_{52}&=8, & c^{(4)}_{53}&=-8, & c^{(4)}_{54}&=0, & c^{(4)}_{55}&=16, \nn
c^{(4)}_{56}&=0, & c^{(4)}_{57}&=0, & c^{(4)}_{58}&=2\pi^2, & c^{(4)}_{59}&=\tfrac{16}{3}, & c^{(4)}_{61}&=0, \nn
c^{(4)}_{62}&=0, & c^{(4)}_{63}&=\tfrac83, & c^{(4)}_{65}&=\tfrac83, & c^{(4)}_{67}&=4, & c^{(4)}_{68}&=0, \nn
c^{(4)}_{70}&=8, & c^{(4)}_{72}&=\tfrac{16}{3}, & c^{(4)}_{74}&=4, & c^{(4)}_{75}&=0, & c^{(4)}_{77}&=4, \nn
c^{(4)}_{80}&=\tfrac{8}{3}, & c^{(4)}_{85}&=0, & c^{(4)}_{86}&=16, & c^{(4)}_{87}&=24, & c^{(4)}_{88}&=8, \nn
c^{(4)}_{89}&=24, & c^{(4)}_{90}&=24, & c^{(4)}_{91}&=\pi^2, & c^{(4)}_{92}&=2\pi^2, & c^{(4)}_{93}&=\pi^2, \nn
c^{(4)}_{94}&=-2, & c^{(4)}_{95}&=8, & c^{(4)}_{97}&=\tfrac{8}{3}, & c^{(4)}_{98}&=2\pi^2, & c^{(4)}_{99}&=2\pi^2, \nn
c^{(4)}_{100}&=2\pi^2, & c^{(4)}_{101}&=2\pi^2, & c^{(4)}_{103}&=\tfrac{\pi^2}{2}, & c^{(4)}_{105}&=\tfrac{\pi^2}{2}, & & 
\label{msbar}
\end{align}
One can therefore see that these coefficients satisfy every equation in Eqs.~(\ref{cc1})--(\ref{cc6}). Evaluating the integrals in Table \ref{fig2} and subtracting the central two-loop subdivergences, we find that
\begin{equation}
c^{(4)}_{69}=\tfrac{4}{27},\;\;\;\; c^{(4)}_{82}=\tfrac{22}{27},
\label{AD}
\end{equation}
and hence using \eqref{ADpredict} the other anomalous dimension coefficients are predicted to be
\begin{align}
c^{(4)}_{60}&=\tfrac49, & c^{(4)}_{64}&=-\tfrac89, & c^{(4)}_{66}&=-\tfrac49, & c^{(4)}_{71}&=\tfrac{16}{9} \nn
c^{(4)}_{73}&=\tfrac49 & c^{(4)}_{76}&=\tfrac89, & c^{(4)}_{78}&=\tfrac{32}{27}, & c^{(4)}_{79}&=\tfrac{20}{27}, \nn
c^{(4)}_{81}&=-\tfrac49, & c^{(4)}_{83}&=\tfrac{\pi^2}{3} & c^{(4)}_{84}&=\tfrac{\pi^2}{3}, & c^{(4)}_{96}&=\tfrac29,\nn
c^{(4)}_{102}&=\tfrac{\pi^2}{3},&c^{(4)}_{104}&=\tfrac{\pi^2}{3}.&&&&
\end{align}
We have checked most, though not all, of these predictions by explicit computation.
  
Finally, a word on scheme dependence. A change in scheme can be effected by a redefinition of the couplings. In our case we may consider
\be
\delta Y^{(2)}=\sum_{\alpha=1}^5\delta_{\alpha}U^{(2)}_{\alpha}
\ee
where the $U^{(2)}_{\alpha}$ are defined in Eq.~\eqref{Ynone} (we continue to focus on the scalar-fermion case and omit potential gauge contributions). The resulting changes in $\beta^{(4)}_{Y}$ and $A^{(7)}$ are given by
\begin{align}
\delta A^{(7)}=&-  \delta Y^{(2)}\cdot \frac{\pa}{\pa{Y}}\;A^{(5)},\nn
\delta \beta^{(4)}_{Y} =&
\beta^{(2)}_{Y}\cdot \frac{\pa}{\pa{Y}} \;\delta Y^{(2)} -  \delta Y^{(2)}\cdot \frac{\pa}{\pa{Y}}\; \beta_{Y}^{(2)}.
\label{betred}
\end{align}
 Using the two-loop $\beta$-function (for the non-gauged theory) as given by Eqs.~\eqref{betdef}, \eqref{Ynone}, \eqref{valstwo}, the induced changes in the $a$-function coefficients as defined in Eq.~\eqref{A7Y} are given by
\begin{align}
\delta a^{(7)}_{5}&=-\tfrac43\delta_5,&
\delta a^{(7)}_{6}&=-\tfrac43\delta_4,&
\delta a^{(7)}_{23}&=-8\delta_1,\nn
\delta a^{(7)}_{24}&=-8\delta_1,&
\delta a^{(7)}_{26}&=-8\delta_1,&
\delta a^{(7)}_{27}&=-8(4\delta_1+\delta_2),\nn
\delta a^{(7)}_{29}&=-\tfrac43\delta_1+16\delta_5,&
\delta a^{(7)}_{30}&=-8\delta_1,&
\delta a^{(7)}_{31}&=-\tfrac43\delta_1+16\delta_5,\nn
\delta a^{(7)}_{32}&=-\tfrac83\delta_2+8\delta_5,&
\delta a^{(7)}_{33}&=-8\delta_2,&
\delta a^{(7)}_{34}&=-\tfrac43\delta_5,\nn
\delta a^{(7)}_{35}&=-\tfrac43\delta_1+16\delta_4,&
\delta a^{(7)}_{36}&=-\tfrac43\delta_1+16\delta_4,&
\delta a^{(7)}_{37}&=-8(\delta_2+\delta_3),\nn
\delta a^{(7)}_{38}&=-\tfrac83\delta_3+8\delta_5,&
\delta a^{(7)}_{39}&=-\tfrac83(\delta_4+\delta_5),&
\delta a^{(7)}_{40}&=-\tfrac83\delta_2+8\delta_4,\nn
\delta a^{(7)}_{41}&=-\tfrac43\delta_4,&
\delta a^{(7)}_{44}&=-8(4\delta_1+\delta_3),&
\delta a^{(7)}_{46}&=-8\delta_3,\nn
\delta a^{(7)}_{47}&=-\tfrac83\delta_3+8\delta_4,&&
\label{aredef}
\end{align}
and the corresponding changes in the $\beta^{(4)}_{Y}$ coefficients defined by are found to be 
\begin{align}
\delta c^{(4)}_{52}&=2(\delta_{1}-4\delta_2) & \delta c^{(4)}_{53}&=2(\delta_{1}-4\delta_2) & \delta c^{(4)}_{54}&=4(4\delta_2-\delta_{1}) & \delta c^{(4)}_{55}&=2(4\delta_2-\delta_{1}) \nn
\delta c^{(4)}_{56}&=4(4\delta_2-\delta_{1}) & \delta c^{(4)}_{57}&=2(\delta_{1}-4\delta_2) & \delta c^{(4)}_{59}&=\tfrac23\delta_{1}-8\delta_4 & \delta c^{(4)}_{60}&=8\delta_4-\tfrac23\delta_{1} \nn
\delta c^{(4)}_{63}&=\tfrac23\delta_{1}-8\delta_4 & \delta c^{(4)}_{64}&=8\delta_4-\tfrac23\delta_{1} & \delta c^{(4)}_{65}&=\tfrac43\delta_2-4\delta_4 & \delta c^{(4)}_{66}&=4\delta_4-\tfrac43\delta_2 \nn
\delta c^{(4)}_{70}&=\tfrac23\delta_{1}-8\delta_5 & \delta c^{(4)}_{71}&=8\delta_5-\tfrac23\delta_{1} & \delta c^{(4)}_{72}&=\tfrac23\delta_{1}-8\delta_5 & \delta c^{(4)}_{73}&=8\delta_5-\tfrac23\delta_{1} \nn
\delta c^{(4)}_{74}&=4(\delta_3-\delta_2) & \delta c^{(4)}_{75}&=16(\delta_2-\delta_3) & \delta c^{(4)}_{76}&=4\delta_4-\tfrac43\delta_3 & \delta c^{(4)}_{77}&=\tfrac43\delta_3-4\delta_4 \nn
\delta c^{(4)}_{78}&=\tfrac43(\delta_4-\delta_5) & \delta c^{(4)}_{79}&=\tfrac43(\delta_5-\delta_4) & \delta c^{(4)}_{80}&=\tfrac43\delta_2-4\delta_5 & \delta c^{(4)}_{81}&=4\delta_5-\tfrac43\delta_2 \nn
\delta c^{(4)}_{85}&=2(4\delta_3-\delta_{1}) & \delta c^{(4)}_{86}&=2(\delta_{1}-4\delta_3) & \delta c^{(4)}_{87}&=2(\delta_{1}-4\delta_3) & \delta c^{(4)}_{88}&=2(\delta_{1}-4\delta_3) \nn
\delta c^{(4)}_{89}&=2(4\delta_3-\delta_{1}) & \delta c^{(4)}_{90}&=2(4\delta_3-\delta_{1}) & \delta c^{(4)}_{96}&=4\delta_5-\tfrac43\delta_3 & \delta c^{(4)}_{97}&=\tfrac43\delta_3-4\delta_5,
\label{deltacs}
\end{align}
all the other coefficients remaining unchanged. Given the method of derivation, it is expected that the consistency conditions will be scheme-independent.
It is indeed easy to verify that all consistency conditions, including the expressions for the anomalous dimension coefficients, are invariant under the changes in Eq,~\eqref{deltacs}, and hence hold in an arbitrary renormalization scheme. This constitutes an additional check on the validity of these consistency conditions. We finally remark that at this order and in the ungauged case, we have from Eqs.~\eqref{Agen}, \eqref{aone} that 
$A^{(5)}=\tfrac14Y_{abij}(\beta_Y^{(2)})_{abij}$ and then the freedom in $A^{(7)}$ corresponds simply to taking 
$\delta Y^{(2)}=-\tfrac14a\beta_Y^{(2)}$. It is in fact easy to check from Eqs.~\eqref{betred}, \eqref{aredef} that this reproduces at lowest order the freedom expressed by the $a$ term in  Eq.~\eqref{Aseven} but leaves $\beta_Y^{(4)}$ unchanged. This redefinition of course vanishes at the fixed point and so the fixed point coupling is unchanged, as is the fixed-point value of the $A$-function. More general coupling redefinitions will correspond to a change in renormalisation scheme with an attendant change in the $\beta$-functions, and furthermore the fixed point coupling value and fixed point $A$-function value will correspondingly be redefined. 

\section{Prediction for general Yukawa-$\beta$-function}
\label{A3}
As we explained in Section 3, if we assume that the $a$-theorem in three dimensions does indeed hold, then 
we may use it to derive a prediction for the part of the four-loop Yukawa $\beta$-function involving factors of the scalar coupling $h$, in the full gauged case.  We now write
 \be
 \beta^{(4)}_{Y}=\sum_{\alpha=1}^{13}c_{H_{\alpha}}U^{(4)}_{H_{\alpha}}+....
\label{bethy}
 \ee
extending the mixed scalar-Yukawa terms in Eq.~\eqref{bydef} to the gauged case.
As explained earlier, the tensor structures $ U^{(4)}_{h_{\alpha}}$,  $\alpha=1\ldots13$, may be read off from the vertices in Table~\ref{fig1}; the ellipsis in Eq.~\eqref{bethy} subsumes all the contributions with no factors of $h$. We computed the pure Yukawa contributions in Sect. 3, but of course there will also be mixed Yukawa-gauge contributions in this general case. The coefficients $c^{(4)}_{H_1}$--$c^{(4)}_{H_6}$ were already given in Eq.~\eqref{cH}.  We now examine the consequences of Eq.~\eqref{dy}. It is easy to see that there is no mixed scalar/Yukawa contribution from the $T^{(5)}_{YY}$ terms, since there is no possible contribution to $T^{(5)}_{YY}$ itself containing $h$ and of course
$\beta^{(2)}_Y$ does not contain $h$ either. Eqs.~\eqref{adef}, \eqref{bethy} now  imply
\begin{align}
c^{(4)}_{H_7}&=\lambda d^{(2)}_{10},& c^{(4)}_{H_8}&=2\lambda d^{(2)}_{10},&
c^{(4)}_{H_9}&=\lambda d^{(2)}_{11},\nn
 c^{(4)}_{H_{10}}&=2\lambda d^{(2)}_{11},&
 c^{(4)}_{H_{11}}&=2\lambda d^{(2)}_{12},& 
   c^{(4)}_{H_{12}}&=2\lambda d^{(2)}_{13},\nn
 c^{(4)}_{H_{13}}&=\lambda d^{(2)}_{14},&&
\label{hycoeffs}\end{align}
with no contribution to $T^{(5)}_{YY}$ in Eq.~\eqref{gradseven}.
 These  coefficients form a prediction for the part of the four-loop Yukawa $\beta$-function involving scalar-coupling contributions; namely, combining Eqs.~\eqref{scalvals}, \eqref{cH}, \eqref{hycoeffs} and taking $\lambda=\tfrac{1}{90}$,
\begin{align}
\beta_Y^{(4)}=&\tfrac13U^{(4)}_{H_1}+\tfrac{2}{45}U^{(4)}_{H_2}
+4(-2U^{(4)}_{H_3}-2U^{(4)}_{H_4}+2U^{(4)}_{H_5}+2U^{(4)}_{H_6}-U^{(4)}_{H_7}-2U^{(4)}_{H_8}
+U^{(4)}_{H_9}+2U^{(4)}_{H_{10}}\nn
&+4U^{(4)}_{H_{13}}+4U^{(4)}_{H_{14}})+\ldots 
\label{pred}
\end{align}
(where we have included the non-gauge terms $U^{(4)}_{H_1}$-- $U^{(4)}_{H_6}$.
This subsumes all the contributions to $\beta_Y^{(4)}$ for a general abelian Chern-Simons theory involving a factor of $h$; as we mentioned, the  purely $Y$-dependent terms were obtained in Sect. 3. Of course the computation of the remaining mixed Yukawa-gauge terms would still require considerable labour, even after exploiting any additional consistency conditions which might arise.

Since there are many potential four-loop Yukawa $\beta$-function structures involving $h$ which are not included in Eq.~\eqref{bethy}, this prediction might appear to give a great deal of additional information in the form of requirements for vanishing coefficients; but we should also consider the relations among these coefficients following from Eq.~\eqref{gauge}. In the case of contributions with  two gauge matrices, the four structures 
$U^{(4)}_{H_7}$--$U^{(4)}_{H_{10}}$ may be extended to a basis by adding just one more structure, whose coefficient in the $\beta$-function is easily seen to be zero without any calculation. Therefore Eq.~\eqref{grad} effectively yields only the four constraints expressed by the coefficient predictions in Eq.~\eqref{hycoeffs}. Similar remarks would be expected to apply to the contributions with four and six gauge matrices.

\end{document}